\documentclass[prd,twocolumn,showpacs,preprintnumbers,floats,floatfix,nofootinbib]{revtex4}

\usepackage{graphicx}
\usepackage{dcolumn}
\usepackage{amsfonts}
\usepackage{amssymb}

\begin{document}

\title{Mass-Energy and Momentum Extraction by Gravitational Wave Emission
in the Merger of Two Colliding Black Holes: The Non-Head-On Case}

\author{R. F. Aranha$^{1,2}$, I. Dami\~ao Soares$^{1}$ and E. V. Tonini$^{3}$}

\address{$^{1}$Centro Brasileiro de Pesquisas F\'isicas, Rio de Janeiro 22290-180, Brazil,\\
$^{2}$Center for Relativistic Astrophysics, Georgia Institute of Technology, Atlanta, GA 30332, USA,\\
$^{3}$Instituto Federal do Esp\'irito Santo, Vit\'oria 29040-780, Brazil.}
\email{rafael.aranha@physics.gatech.edu;\\ ivano@cbpf.br;tonini@cefetes.br}


\begin{abstract}

We examine numerically the post-merger regime of two Schwarzschild black holes in non head-on collision.
Our treatment is made in the realm of non-axisymmetric Robinson-Trautman spacetimes which are appropriate
for the description of the system. Characteristic initial data for the system are constructed and
the Robinson-Trautman equation is integrated for these data using a numerical code based on the Galerkin spectral method
which is accurate and sufficiently stable to reach the final configuration of the remnant black hole,
when the gravitational wave emission ceases. The initial data contains three independent parameters,
the ratio mass $\alpha$ of the individual colliding black holes, the boost parameter (that characterizes the initial pre-merger
infalling velocity of the two black holes) and the incidence angle of collision $0 \leq \rho_0 \leq 90^{o}$. The remnant black hole is
characterized by its final boost parameter, the final rest mass and scattering angle.
The motion of the remnant black hole is restricted to the plane determined by the directions of the two initial colliding
black holes, characterizing a planar collision. The energy-momentum fluxes carried out by gravitational waves
are confined to this plane. We evaluate the efficiency of mass-energy extraction, the total energy and momentum carried out
by gravitational waves and the momentum distribution of the remnant black hole for a large domain of initial data parameters.
Our analysis is based on the Bondi-Sachs four momentum conservation laws. The process of mass-energy extraction is shown to be less efficient
as the initial data departs from the head-on configuration. Head-on collisions ($\rho_0=0$) and orthogonal
collisions ($\rho_0=90^{o}$) constitute, respectively, upper and lower bounds to the power emission and
to the efficiency of mass-energy extraction. On the contrary, head-on collisions and orthogonal collisions constitute,
respectively, lower and upper bounds for the momentum of the remnant. The momentum extraction and the pattern of the
momentum fluxes, as a function of the incidence angle, are examined. The momentum extraction characterizes a regime of
strong deceleration of the system. The angular pattern of gravitational wave signals are also examined for small mass
ratios $\alpha$ and early times $u$. We show that, in the plane of collision $(x,z)$, the pattern is typically bremsstrahlung,
corresponding to a strong deceleration regime at early times, with two dominant lobes in the forward direction of motion.
Gravitational waves are also emitted outside the plane of collision, the angular distribution of which is symmetric with respect to the $y$-axis
and therefore with a zero net momentum, consistent with the planar nature of the collision.
Finally the relation between the incidence angle and the scattering angle closely approximates a relation
for the inelastic collision of classical particles in Newtonian dynamics.

\end{abstract}

\maketitle

\section{Introduction}

It is by now theoretically well established that, in the nonlinear regime of
General Relativity, gravitational waves extract mass, momentum and angular momentum of the
source and that the radiative transfer involved in these processes may turn out to
be fundamental for the astrophysics of the collapse of stars, the formation of black holes
and the collision and merger of two or more black holes. In the present paper we will examine
the radiative transfer processes by gravitational wave emission in the post-merger
phase of two colliding black holes, with focus on the case of non-head-on collisions,
described in the realm of Robinson-Trautman (RT) spacetimes\cite{rt}. This work extends two
previous recent papers\cite{aranha1,aranha2} where we examined the case of head-on collisions.
The collision and merger of two black holes are considered to be important astrophysical
sources of strong gravitational emission (cf.\cite{pretorius} and references therein) and the
relevance of the processes of generation and emission of gravitational waves in these configurations
lies in the fact that the associated wave patterns will be crucial for the present efforts towards a direct
detection of gravitational waves. In spite of the enormous progress achieved until now
using approximation methods and numerical techniques, the information on wave form patterns
and radiative transfer processes in the dynamics of gravitational wave emission is far from
being complete\cite{boyle}.
\par The object of this paper is the numerical treatment of the gravitational wave production
and the associated processes of radiative transfer in the non-head-on collision
of two boosted Schwarzschild black holes, modeled in the context of Robinson-Trautman (RT)
spacetimes. Characteristic initial data for the RT dynamics are constructed that represent
instantaneously two black holes in non-head-on collision. RT spacetimes present a global apparent
horizon\cite{AH} so that the dynamics corresponds to a regime where the merger has already set in.
During the merger part of the rest mass and kinetic energy of the
two initial individual black holes are radiated away by the gravitational waves, and part
will be absorbed to constitute the mass-energy of the remnant. The linear momentum of the
initial system is also carried out by the gravitational waves emitted. We also examine the
two regimes of momentum extraction in the post-merger phase by contemplating the time behavior of the
momentum flux of the gravitational waves emitted. We use the momentum-energy conservation law in the
Bondi-Sachs formulation of gravitational wave emission by bounded sources\cite{bondi,sachs}.
Also the analysis of the energy flux carried out by gravitational waves will allow us to characterize
two distinct regimes of bremsstrahlung emission for distinct domains of the initial data parameters.
Several characteristics of a non-head-on collision are drastically distinct from a head-on collision,
mainly for large values of the initial mass ratio parameter. The general outcome will be a remnant boosted black hole
with a rest mass larger than the sum of the rest masses of the two individual initial black holes.
\par As compared to previous estimates of the literature, our approach differs basically in that we have adopted the
characteristic surface initial data formalism, which has several advantages for the description
of gravitational radiation and the construction of algorithms\cite{winicour}. In this direction, an accurate
code based on the Galerkin method was constructed to integrate the RT non-axisymmetric field equations.
The code is accurate and highly stable for long time runs in the nonlinear regime so that we
are able to reach numerically the final configuration of the system, when the
gravitational emission ceases. This will allow us to describe processes of mass-energy,
and momentum extraction due to gravitational radiation emission, by using  physically meaningful
quantities connected to initial and final configurations of the source.
\par We organize the paper as follows. In Sec. II we review some basic aspects of
the general non-axisymmetric Robinson-Trautman spacetimes necessary for our analysis of gravitational
radiation emission in the case of a non-head-on collision. In Section III we construct characteristic
initial data for the non-axisymmetric RT dynamics corresponding to two boosted Schwarzchild black holes in a non-head-on collision.
In Section IV we present a basic description of the numerical codes used to evolve the characteristic initial data via
the RT dynamics. The codes are based on Galerkin spectral methods and the dynamical evolution of the initial data is performed and discussed,
including its accuracy and stability. In Section V we discuss the planar nature of the collision and show how this
property allows us to save a lot of computational effort. In Section VI the Bondi-Sachs
four momentum for the non-axisymmetric RT spacetimes are introduced together with its conservation laws
that will be fundamental in the treatment of the energy and linear momentum extracted from the source by the gravitational waves emitted.
Sections VII to X contain the main results of the paper, as the numerical evaluation of the
efficiency of mass-energy transfer by gravitational wave emission, the conservation of energy in the processes via the Bondi
mass formula and linear momentum extraction by gravitational waves. Distinct patterns of gravitational wave emission
are discussed related to the range of the physical parameters of the initial data. In Section XI we summarize our results
and discuss their relevance and limitations as compared to previous results in the literature.
Throughout the paper we use units such that $8 \pi G=c=1$.
\section{Dynamics of Robinson-Trautman Spacetimes. Wave Zone Curvature}

RT spacetimes\cite{rt} are asymptotically flat solutions of Einstein's vacuum equations
that describe the exterior gravitational field of a bounded system
radiating gravitational waves. The RT metric is expressed as
{\small
\begin{eqnarray}
\label{eq1}
\nonumber
ds^2=\alpha^2(u,r,\theta,\phi) d u^2+2du dr -r^{2}K^{2}(u,\theta,\phi)\\
\times (d \theta^{2}+\sin^{2}\theta d \varphi^{2}).
\end{eqnarray}}
where $r$ is an affine parameter defined along the shearfree
null geodesics determined by the vector field $\partial/\partial r$. Einstein equations imply that,
in a suitable coordinate system,
{\small
\begin{eqnarray}
\label{eq2}
\alpha ^{2} (u,r,\theta,\phi)= \lambda(u,\theta,\phi)- \frac{2 m_{0}}{r}+ 2 r \frac{{K}_u}{K} ,
\end{eqnarray}}
where $m_0$ is a positive constant, and  $\lambda(u,\theta,\phi)$ is the
Gaussian curvature of the surfaces (u={\rm const}, r={\rm const}) defined by
{\small
\begin{eqnarray}
\lambda=\frac{1}{K^2}-\frac{(K_{\theta}~ \sin \theta/K)_{\theta}}{K^2 \sin \theta}
+\frac{1}{\sin^2 \theta}\Big(\frac{K_{\phi}^2}{K^4}-\frac{K_{\phi \phi}}{K^3} \Big).
\label{eq3}
\end{eqnarray}}
The remaining Einstein equations yield
{\small
\begin{eqnarray}
\label{eq4} -6 m_{0}\frac{{K}_u}{K}+\frac{1}{2 K^2}\Big(\frac{(\lambda_{\theta} \sin
\theta)_{\theta}}{ \sin \theta}+\frac{\lambda_{\phi \phi}}{ \sin^2 \theta}\Big)=0.
\end{eqnarray}}
\noindent In the above, subscripts $u$, $\theta$ and $\phi$ denote derivatives with respect to
$u$, $\theta$ and $\phi$, respectively. Eq. (\ref{eq4}), denoted RT equation, governs the dynamics of the
gravitational field which is totally contained in the metric function $K(u,\theta,\phi)$.
Chrusciel and Singleton\cite{chrusciel} established that RT spacetimes exist globally for all positive
$u$ and converge asymptotically to the Schwarzschild metric as $u \rightarrow \infty$
-- this global time extension being realized for arbitrary smooth initial data.
\par An important feature of RT spacetimes, that establishes its radiative
character, arises from the expression of its curvature tensor that in the semi-null tetrad basis
{\small
\begin{eqnarray}
\label{eq5}
\nonumber
\Theta^{0}&=&du,~~~~~~~ \Theta^{1}=(\alpha^{2}/2)~ du+dr\\
\Theta^{2}&=&r K d \theta,~~~ \Theta^{3}=r K \sin \theta d\phi
\end{eqnarray}}
assumes the form
{\small
\begin{eqnarray}
\label{eq6}
R_{ABCD}=\frac{N_{ABCD}}{r}+\frac{III_{ABCD}}{r^2}+\frac{II_{ABCD}}{r^3},
\end{eqnarray}}
where the scalar quantities {\small $N_{ABCD}$, $III_{ABCD}$ and $II_{ABCD}$} are of the
algebraic type $N$, $III$ and $II$, respectively, in the Petrov classification
of the curvature tensor\cite{petrov}, and $r$ is the parameter distance along the
principal null direction $\partial /\partial r$. Eq. (\ref{eq6}) displays the peeling
property\cite{peeling} of the curvature tensor, showing that indeed RT is
the exterior gravitational field of a bounded source emitting gravitational waves.
For large $r$ we have
{\small
\begin{equation}
R_{ABCD} \sim \frac{N_{ABCD}}{r},
\label{eq7}
\end{equation}}
so that at large $r$ the gravitational field looks like a gravitational wave with propagation
vector $\partial /\partial r$. The nonvanishing of the scalars {\small $N_{ABCD}$}
is an invariant criterion for the presence of gravitational waves, and the
asymptotic region where ${\cal{O}}(1/r)$-terms are dominant defined as the wave zone.
The curvature tensor components in the above basis that contribute to {\small $N_{ABCD}$}, namely, to
the gravitational degrees of freedom transversal to the direction of propagation of the wave, are
{\small $R_{0303}=-R_{0202}=-D(u,\theta,\phi)/r+{\cal{O}}(1/r^2)$} and {\small $R_{0203}=-B(u,\theta,\phi)/r+{\cal{O}}(1/r^2)$} where
{\small
\begin{eqnarray}
\label{eq8}
\nonumber
D(u,\theta,\phi)=\frac{1}{2K^2} ~\partial_{u}\Big(
\frac{K_{\theta \theta}}{K}-\frac{K_{\theta}}{K} \cot \theta-
\frac{2~K_{\theta}^2}{K^2}\Big)\\
+\frac{1}{2 K^2 \sin^2 \theta}~\partial_{u} \Big(-\frac{K_{\phi \phi}}{K}+\frac{2 K_{\phi}^{2}}{K^2} \Big),
\end{eqnarray}}
and
{\small
\begin{eqnarray}
\label{eq9} B(u,\theta,\phi)=\frac{1}{K^2 \sin \theta}~ \partial_{u}
\Big(\frac{K_{\theta \phi}}{K}-2\frac{K_{\theta} K_{\phi}}{K^2} -\cot \theta \frac{K_{\phi}}{K}\Big),
\end{eqnarray}}
\noindent From (\ref{eq7}) we can see that the functions $D$ and $B$ contain
all the information of the angular, and time dependence of the gravitational
wave amplitudes in the wave zone once $K(u,\theta,\phi)$ is given. $D$ and $B$
actually correspond to the two independent polarization modes of the gravitational wave, transverse to its direction of
propagation at the wave zone.
\par The field equations present two stationary solutions, which will play a crucial role in our
future discussions. The first is the Schwarzschild solution corresponding to
{\small
\begin{equation}
K=K_0= {\rm const},~~~~~\lambda=1/K_{0}^2
\label{eq10}
\end{equation}}
and mass $M_{Schw}=m_{0} K_{0}^{3}$.
The second is
{\small
\begin{equation}
K(\theta,\phi)=\frac{K_0}{\cosh \gamma+ ({\bf{n}} \cdot \hat{\bf{x}})\sinh \gamma},
\label{eq11}
\end{equation}}

\noindent where $\hat{\bf{x}}=(\sin \theta \cos \phi,\sin \theta \sin \phi,cos \theta)$ is the unit
vector along an arbitrary direction ${\bf{x}}$ and ${\bf{n}}=(n_1,n_2,n_3)$ is a constant unit vector
(satisfying $n_{1}^{2}+n_{2}^{2}+n_{3}^{2}=1$); also $K_0$ and $\gamma$ are constants. We note
that (\ref{eq11}) yields $\lambda=1/K_{0}^{2}$, resulting in its stationary character.
This solution can be interpreted\cite{bondi} as a boosted black hole along the axis determined by the unit vector
${\bf{n}}$ with boost parameter $\gamma$, or equivalently, with velocity parameter $v=\tanh \gamma$.
The $K(\theta,\phi)$ function (\ref{eq11}), which depends on three parameters, is a K-transformation of the generalized
Bondi-Metzner group\cite{bondi} discussed by Sachs\cite{sachs1} (the BMS group) and represents the
general form of Lorentz boosts contained in the homogeneous orthochronous Lorentz transformations of the BMS.
\par The Bondi mass function of this solution is given by $m(\theta,\phi)=m_{0}K^3(\theta,\phi)$. The total
mass-energy of this gravitational configuration is given by the Bondi mass
{\small
\begin{eqnarray}
\label{eq12}
\nonumber
M&=&(1/4\pi)\int^{2\pi}_{0} d \phi \int^{\pi}_{0} d\theta~ m(\theta,\phi)\sin \theta\\
&=&m_{0} K_{0}^{3}\cosh \gamma = m_{0} K_{0}^{3}/ \sqrt{1-v^2}.
\end{eqnarray}}
The interpretation of (\ref{eq11}) as a boosted black hole is relative to
the asymptotic Lorentz frame which is the rest frame of the black hole when $\gamma=0$.
\section{Characteristic Initial Data}

\par As well known the initial data problem for RT spacetimes
is within the class of characteristic initial value formulation as opposed
to the 1+3 formulation, according to the classification of York\cite{york}.
For RT spacetimes the function $K(u_0,\theta,\phi)$ given in a characteristic surface
$u=u_0$ corresponds to the initial data to be evolved via the RT equation (\ref{eq4}).
Our task is now to exhibit an initial $K(u_0,\theta,\phi)$ that represents instantaneously
the general collision of two Schwarzschild black holes (head-on collision or not),
by extending a procedure outlined in Refs. \cite{aranha1,aranha3} to construct
initial data for RT dynamics in the non-axial case.
\par In analogy to bispherical coordinates\cite{arfken} in the 3-dim Cartesian plane
$\Sigma$, let us introduce the following parametrization for Cartesian coordinates
{\small
\begin{eqnarray}
\label{eq3R}
\nonumber
x&=& \frac{a ~\sin \theta~ \sinh \eta}{\cosh \eta + \cos \theta~ \sinh{\eta}}~ \cos \phi,\\
y&=& \frac{a~ \sin \theta~ \sinh \eta}{\cosh \eta + \cos \theta~ \sinh{\eta}}~ \sin \phi,\\
\nonumber
z&=& \pm \frac{a}{\cosh \eta + \cos \theta~ \sinh{\eta}},
\end{eqnarray}
}
for $z>0$ and $z<0$ respectively. In the above $0 \leq \eta \leq \infty$, $0 \leq \theta \leq \pi$,
$0 \leq \phi \leq 2 \pi$. In this parametrization, $\eta=\eta_0$ corresponds to two spheres, one
at $z>0$ and the other at $z<0$, centered at ($x=y=0,z= \pm a \cosh \eta_0$) respectively, with radius
$a \sinh \eta_0$. The Cartesian vector from a point $P: (x,y,z)$ of $\Sigma$ has length
{\small
\begin{eqnarray}
r(\eta,\theta)=a \sqrt{\frac{\cosh \eta - \cos \theta \sinh \eta}{\cosh \eta
+ \cos \theta \sinh \eta}}
\label{eq3_3}
\end{eqnarray}
}
\noindent For $\eta=\infty$ the spheres degenerate into the planes
$z=0$ and $z= \pm \infty$. The usefulness of this parametrization will become clear in what follows. We
note that the Cartesian coordinates are continuous functions, with continuous derivatives, of ($\eta,\theta,\varphi$).
Singularities occurring are the usual singularities of a spherical coordinate system. For future reference let us
introduce the functions
\begin{eqnarray}
{\cal{S}}_{(\pm)}(\eta,\theta,\phi,{\bf{n}})=\sqrt{\cosh \eta \pm
({\bf{n}} \cdot \hat{\bf{x}})\sinh \eta}~.
\label{eq3_4-0}
\end{eqnarray}
where $\hat{\bf{x}}=(\sin \theta \cos \phi,\sin \theta \sin \phi, \cos \theta)$ and ${\bf{n}}=(n_{1},n_{2},n_{3})$,
with $n_{1}^{2}+n_{2}^{2}+n_{3}^{2}=1$. In the above parametrization (\ref{eq3R}), the flat space line element
$ds^2=(dx)^2+(dy)^2+(dz)^2$ is expressed as
{\small
\begin{eqnarray}
ds^2_{\rm flat}=\frac{a^2}{{\cal{S}}_{(+)}^{4}(\eta,\theta,\phi,{\bf{n}})}\Big[d
\eta^2 + \sinh^2{\eta}~(d\theta^2+\sin^2 \theta d \phi^2) \Big].~~
\label{eq3i}
\end{eqnarray}}
\par We now take $\Sigma$ as a spacelike surface of initial data, with geometry
defined by the line element
{\begin{widetext}
{\small
\begin{eqnarray}
ds^2=a^2 K^2(\eta+\gamma_0,\theta,\phi)~\Big[d \eta^2
+\sinh^2({\eta+\gamma_0})(d\theta^2+\sin^2 \theta d \phi^2) \Big]~
\label{eq3ii}
\end{eqnarray}}
\end{widetext}}

\noindent where $\gamma_0$ is an arbitrary parameter. By assuming time-symmetric data (namely, $\Sigma$ a
maximal slice with zero extrinsic curvature) we obtain that the Hamiltonian constraints reduce to
$^{(3)}R=0$. With the substitution $K \equiv \Phi^2$, the constraint equation reduces to the Laplace equation
{\small
\begin{eqnarray}
\nonumber
\frac{1}{\sin \theta}\Big( \Phi_{\theta}\sin \theta \Big)_{\theta}
+\Big(\Phi^{\prime} \sinh^2(\eta+\gamma_0) \Big)^{\prime}\\
+\frac{1}{\sin^2 \theta}~ \Phi_{\phi \phi}+\frac{3}{4}\sinh^2(\eta+\gamma_0) \Phi=0
\label{eq3_8},
\end{eqnarray}}
\noindent where a prime denotes derivative with respect to $\eta$.
It is not difficult to verify that the functions
{\small
\begin{eqnarray}
\Phi=\left(\frac{1}{{\cal{S}}_{(\pm)}(\eta+\gamma_0,\theta,\phi,{\bf{n}})}\right)
\label{eq3_9}
\end{eqnarray}}
satisfy Eq. (\ref{eq3_8}) and, with respect to metric (\ref{eq3ii}),
correspond to flat space solutions (zero curvature). It then follows that
{\small
\begin{eqnarray}
\Phi=\left(\frac{\alpha_1}{{\cal{S}}_{(-)}(\eta+\gamma_0,\theta,\phi,{\bf{n}})}\right)+
\left(\frac{\alpha_2}{{\cal{S}}_{(+)}(\eta+\gamma_0,\theta,\phi,{\tilde{{\bf{n}}}})}\right)
\label{eq3_10}
\end{eqnarray}}
is a nonflat solution of (\ref{eq3_8}), where $\alpha_1$ and
$\alpha_2$ are arbitrary positive constants. The nonflat 3-dim geometry defined by (\ref{eq3_10}),
{\small
\begin{eqnarray}
ds^2= a^2~\Phi^4~\Big[d \eta^2
+\sinh^2{(\eta+\gamma_0)}(d\theta^2+\sin^2 \theta d \phi^2) \Big],
\label{eq3_11}
\end{eqnarray}}

\noindent is asymptotically flat with a form analogous to that of the 3-dim spatial
section of the Schwarzschild geometry in isotropic coordinates, as we proceed to show.
\par Without loss of generality we take in (\ref{eq3_10}) ${\tilde{{\bf{n}}}}=(0,0,1)$, that corresponds
to choose the $z$-axis along ${\tilde{{\bf{n}}}}$. In this instance a straightforward manipulation
shows that, for $\eta >> \gamma_0$, the metric (\ref{eq3_11}) can be rewritten as
{\small
\begin{eqnarray}
ds^2=\left({\alpha_2}+\frac{a{\alpha_1}}{r(\eta,\theta)}~\frac{\sqrt{\cosh \eta - \cos \theta \sinh \eta}}{\sqrt{\cosh \eta - ({\bf{n}}\cdot \hat{\bf{x}}) \sinh \eta}} \right)^4ds^{2}_{{\rm flat}}.~~~
\label{eq3_11i}
\end{eqnarray}}
\par Now to probe the asymptotic structure of the metric (\ref{eq3_11i})
let us consider $\eta$ very large and, for this $\eta$, points
($x,y,z$) whose distance from the origin is also very large, namely,
when ($\eta \rightarrow \infty,\theta \simeq \pi$). In this asymptotic limit,
returning to Cartesian coordinates, the 3-geometry (\ref{eq3_11i}) can be
given in the approximate form
{\small
\begin{eqnarray}
g_{ij} \simeq
\Big \{1+ \frac{2~M_{(1)}}{r(\eta,\theta)}\Big\}~\delta_{ij},~~
\label{eq3_12}
\end{eqnarray}}
where we fixed the scale of bispherical-type coordinates by taking $2 {\sqrt {2}} a= m_0 (\alpha_1+\alpha_2) {\sqrt {1+n_3}} ~\alpha_2/\alpha_1$.
The Schwarzschild mass $M_{(1)}=m_0({\alpha_1+\alpha_2})$.
%
\par From the above construction we can now extract initial data for the
RT dynamics, which has its initial value problem on null cones.
Based on the initial data formulation on characteristic
surfaces proposed by D'Inverno and Stachel\cite{stachel,stachel1} --
in which the degrees of freedom of the vacuum gravitational field are
contained in the conformal structure of 2-spheres embedded in a
3-spacelike surface -- we are then led to adopt the conformal structure given
by the conformal factor (\ref{eq3_10}) defined on the surface $\eta=0$,
{\small
\begin{eqnarray}
K(u_0,\theta,\phi)=\Big(\frac{\alpha_1}{{\cal{S}}_{(-)}(\gamma_0,\theta,\phi,{\bf{n}})}+
\frac{\alpha_2}{{\cal{S}}_{(+)}(\gamma_0,\theta,\phi,{\tilde{{\bf{n}}}})}\Big)^2,
\label{eq3_13}
\end{eqnarray}}
\noindent with ${\tilde{\bf{n}}}=(0,0,1)$, as initial data for the RT dynamics. This conformal structure
is to be extended along null bicharacteristics and propagated
along a timelike congruence of the spacetime via RT dynamics.
A restricted space–time may then be constructed locally as the product of the
two-sphere geometry times a timelike plane ($u,{\tilde{r}}$) generated
by a null vector $\partial/\partial {\tilde{r}}$ and a
timelike vector $\partial/\partial u$ with geometry
$d\sigma^2= \alpha^2(u,{\tilde{r}},\theta,\phi) d u^2+ 2 du d {\tilde{r}}$. The four geometry
is then taken as
\begin{eqnarray}
\label{rtX}
\nonumber
ds^2&=& \alpha^2(u,{\tilde{r}},\theta,\phi) d u^2
+ 2 du d {\tilde{r}}\\ &-& {\tilde{r}}^2 K^2(u,\theta,\phi)\Big(d \theta^{2}+\sin^{2}\theta d
\phi^{2}\Big).
\end{eqnarray}
Eq. (\ref{rtX}) is the RT metric, the dynamics of which (ruled by Einstein's vacuum
field equations) propagates the initial data (\ref{eq3_13}) forward in time from
the characteristic initial surface $u = u_0$. We note that Einstein's vacuum equations
demand that the function $\alpha^2(u,{\tilde{r}},\theta,\phi)$ has the form given in (\ref{eq2}).
\par The interpretation of the asymptotically flat initial data (\ref{eq3_13}) as
two instantaneously interacting Schwarzschild black holes boosted along the
$z$-axis is now discussed, based on perturbations of the RT metric (\ref{rtX})
constructed with such data. As shown in Section II, for $\alpha_1=0$ the data correspond in (\ref{rtX}) to a
static Schwarzschild black hole (with the total Bondi mass $m_0 (\alpha_{2})^{6} \cosh \gamma_0$)
boosted along the direction defined by the unit vector ${\tilde{{\bf{n}}}}$, with $v = \tanh \gamma_0$.
For $\alpha_1 \neq 0$, with $\alpha_1<<\alpha_2$, the configuration is no longer static
and cannot therefore be a black hole, but can still be interpreted
as an initially perturbed boosted Schwarzschild black hole.
Conversely the same consideration holds for ($\alpha_1 \neq 0, \alpha_2=0$)
and $\alpha_1 \neq 0$ with $\alpha_2 << \alpha_1$, the latter case corresponding also to
an initially perturbed  boosted Schwarzschild black hole. In this sense we associate the perturbation
with a black hole of relative small rest mass, boosted along the direction ${\bf{n}}$ (or ${\tilde{{\bf{n}}}}$)
in non head-on collision with a larger black hole boosted along the direction ${\tilde{{\bf{n}}}}$ (or ${\bf{n}}$).
The initial infalling velocity of each black hole considered individually is given by $v=\tanh \gamma_0$.
\par Without loss of generality, in the remaining of the paper we fix $\alpha_2=1$ and drop subscripts $1$ and $0$ of
the parameters $\alpha_1$ and $\gamma_0$, respectively, in order to avoid overcluttering in formulae and Figures.
In this instance, the initial data (\ref{eq3_13}) will in principle contain four independent parameters, namely,
($\alpha,\gamma, {\bf{n}}$) with $n_1^2+n_2^2+n_3^2=1$, and assumes the form

{\begin{widetext}
{\small
\begin{eqnarray}
K(u_0,\theta,\phi)=\Big(\frac{1}{\sqrt{{\cosh \gamma+ \cos \theta \sinh \gamma}}}+
\frac{\alpha}{\sqrt{{\cosh \gamma - ({\bf{n}}\cdot \hat{\bf{x}})\sinh \gamma}}}\Big)^2~.
\label{eq3_13-i}
\end{eqnarray}}
\end{widetext}}
We must comment that, since only two black holes are involved in the collision, the data (\ref{eq3_13-i})
should actually depend (besides $\alpha$ and $\gamma$) only on the incidence angle $\rho_0$ between the two initial black holes,
determined by the scalar product ${\bf{n_z}} \cdot {\bf n}$. Therefore we should expect that (i) the resulting dynamics of the system
would not be altered under a rigid rotation of the two initial black holes about the $z$-axis, which could be used
to locate ${\bf n}$ of the initial data in the $x-z$ plane; and consequently (ii) the initial data will result in a planar dynamics
of the collision, a fact that would save a lot of computational effort. The above statements (i)-(ii) will indeed be confirmed
by the numerical results obtained from the evolution of the data (\ref{eq3_13-i}), as shown in section VI, and establish
the planar nature of a general non head-on collision of two black holes.
\par Finally we should remark that, in the full Bondi-Sachs problem, the analysis of field equations in the
$2+2$ formulation\cite{bondi,stachel} shows that the two {\it news} functions $c_{u}^{(1)}(u,\theta,\phi)$ and $c_{u}^{(2)}(u,\theta,\phi)$
are part of the initial data to be prescribed for the evolution of the system\cite{aranha4}. However for the RT dynamics the {\it news} are already
specified once the initial data for the RT equation, namely $K(u_0,\theta,\phi)$, is given and consequently $K(u,\theta,\phi)$ is
given for all $u > u_0$ from the numerical evolution of the data.
\par The initial data (\ref{eq3_13-i}) will be evolved numerically -- via RT dynamics -- up to a final configuration
that corresponds to a remnant Schwarzschild black hole boosted along the axis determined by the unit
vector ${\bf{n}}_f$ and having the form of the solution (\ref{eq11}), as we discuss in the next Section.
A numerical code using the Galerkin spectral method is was implemented to integrate the
non-axisymmetric RT equation (\ref{eq4}), the basis of which is described in the next Section.

\section{Numerical evolution of the data}

We proceed now to discuss the numerical evolution of the initial data (\ref{eq3_13-i}) via the
non-axisymmetric RT equation (\ref{eq4}). Throughout the present Section the variable $\theta$ will be expressed
in terms of the variable $x= \cos \theta$. The numerical integration of the non-axisymmetric RT equation is
performed using a Galerkin spectral method which is now described in detail.
In the present procedure we rewrite the equations (\ref{eq3})-(\ref{eq4}) using the variable
$P(u,x,\phi) \equiv 1/K(u,x,\phi)$ (instead of the former $K(u,x,\phi)$),
an approach already adopted in Ref. \cite{saa}. We obtain that
{\small
\begin{eqnarray}
\label{lambdaP}
\nonumber
\lambda(u,x,\phi)=(1-x^2)\left[PP_{xx}-P_{x}^2\right]-2xPP_{x}+P^2+\\
+\frac{1}{(1-x^2)}\left[PP_{\phi\phi}-P_{\phi}^2\right]
\end{eqnarray}}
\noindent and
{\small
\begin{eqnarray}
\label{Pdot}
{\dot P}(u,x,\phi)=-\frac{P^3}{12m_0}\Big[(1-x^2) \lambda_{xx}-2x\lambda_{x}+\frac{\lambda_{\phi\phi}}{(1-x^2)}\Big].
\end{eqnarray}}
where a dot and the the subscripts $x$ and $\phi$ denote, respectively, derivatives with respect to $u$, $x$ and $\phi$.
\par The Galerkin method establishes that $P(u,x,\phi)$ can be expanded in a convenient set of basis
functions of a projection space by which we can reduce the RT partial differential equation (\ref{Pdot}) into a finite set
of nonlinear coupled ordinary differential equations. This set of equations constitutes an autonomous dynamical system, the dimension of which
depends directly on the truncation in the Galerkin method to approximate the RT dynamics. As the conformal function $P(u,x,\phi)$
defined on the 2-sphere is assumed to be sufficiently smooth, we can use the real Spherical Harmonics (SHs) \cite{korn} as
the appropriate basis that better approximates our desired solution. Before starting with the numerical scheme,
we first present some definitions and properties of these functions.

\par Unlike the complex SHs, the real SHs consist of a break of the complex SHs into their sine and cosine parts as follows,
{\small
\begin{eqnarray}
\nonumber
\left<x,\phi|l,m\right>^{{\small(+)}} &\equiv& ^{{\small{(+)}}}Y_{l}^{m}(x,\phi)\equiv W_{l,m}P_{l}^{m}(x)\cos(m\phi)\\
\nonumber
&=&^{{\small(+)}}\left<l,m|x,\phi\right>,\\
\nonumber
\left<x,\phi|l,m\right>^{{\small(-)}} &\equiv& ^{{\small(-)}}Y_{l}^{m}(x,\phi)\equiv W_{l,m}P_{l}^{m}(x)\sin(m\phi)\\
&=&^{{\small(-)}}\left<l,m|x,\phi\right>,
\end{eqnarray}}
\noindent where $P_{l}^{m}(x)$ are the associated Legendre functions defined by
{\small
\begin{eqnarray}
\label{LegendreA}
P_{l}^{m}(x)\equiv(1-x^2)^{|m|/2}\frac{d^mP_{l}}{dx^m}(x),
\end{eqnarray}}
\noindent and the normalization factors $W_{l,m}$ are given by
{\small
\begin{eqnarray}
\label{normW}
W_{l,m}\equiv \sqrt{\frac{1}{2}\frac{(2l+1)(l-m)!}{\pi (l+m)!}}.
\end{eqnarray}}
\noindent From these formulas we see that the case $m=0$ gives us the basis functions for the axial case,
the Legendre polynomials $P_{l}(x)\equiv\left<x|l\right>$ \cite{korn}.
\par The real SH orthogonality relations are given by
{\small
\begin{eqnarray}
\label{ortho}
\nonumber
&&{^{\small(\pm)}}\left<l,m|l',m'\right>{^{\small(\pm)}} \equiv \\
\nonumber
&\equiv&\int_{0}^{2\pi}\int_{-1}^{1}{^{\small(\pm)}}\left<l,m|x,\phi\right>\left<x,\phi|l',m'\right>{^{\small(\pm)}}
\nonumber
dx~d\phi\\
&=&\int_{0}^{2\pi}\int_{-1}^{1}{^{\small(\pm)}}Y_{l}^{m}(x,\phi){^{\small(\pm)}}Y_{l'}^{m'}(x,\phi)
\nonumber
dx~ d\phi\\
&=&\delta_{l,l'}\delta_{m,m'},~~
\end{eqnarray}
and
\begin{eqnarray}
\label{ortho1}
\nonumber
&& {^{\small(\pm)}}\left<l,m|l',m'\right>{^{\small(\mp)}} \equiv\\
\nonumber
&=&\int_{0}^{2\pi}\int_{-1}^{1}{^{\small(\pm)}}\left<l,m|x,\phi\right>\left<x,\phi|l',m'\right>{^{\small(\mp)}}
dx~d\phi\\
&=&\int_{0}^{2\pi}\int_{-1}^{1}{^{\small(\pm)}}Y_{l}^{m}(x,\phi){^{\small(\mp)}}Y_{l'}^{m'}(x,\phi)
dx~d\phi
=0,~
\end{eqnarray}}
\noindent and are fundamental in the treatment of the Galerkin method. Here the following completeness relation is also used,
{\small
\begin{eqnarray}
\label{identityInt}
\int_{0}^{2\pi}\int_{-1}^{1} {\left|x,\phi\left>\right<x,\phi\right|}~dx~d\phi=1.
\end{eqnarray}}
\par Now we are able to construct our numerical Galerkin scheme, that will allow us to obtain an accurate approximated numerical solution for (\ref{Pdot}) corresponding to given initial data. As every twice continuously differentiable, suitably periodic real function
defined on the surface of a sphere admits an absolutely convergent expansion in terms of the SHs, let us consider the expansion
{\small
\begin{eqnarray}
\label{Pappr}
\nonumber
P_{(a)}(u,x,\phi)=\sum_{l=0}^{N_{P}} \frac{1}{2} A_{l,0}(u) {^{\small(+)}} Y_{l}^{0}(x,\phi)+\\
\nonumber
+\sum_{l=1}^{N_{P}}\Big[ \sum_{m=1}^{l} A_{l,m}(u){^{\small(+)}} Y_{l}^{m}(x,\phi)+\\
+\sum_{m=1}^{l} B_{l,m}(u){^{\small(-)}} Y_{l}^{m}(x,\phi)\Big],
\end{eqnarray}}
\noindent as our approximated solution for $P(u,x,\phi)$. Here $N_P$ is a positive integer that defines the truncation order
of the Galerkin method. By using the orthogonality relations (\ref{ortho}) and (\ref{ortho1}), the modal coefficients
$A_{l,m}(u)$ and $B_{l,m}(u)$ are given by
{\small
\begin{eqnarray}
\label{modal}
\nonumber
A_{l,m}(u)&=&\int_{0}^{2\pi} \int_{-1}^{1} {^{\small(+)}} Y_{l}^{m}(x,\phi) P_{(a)}(u,x,\phi) dx d\phi\\
&\equiv&{^{\small(+)}}\left<l,m|P_{(a)}\right>,\\
\nonumber
B_{l,m}(u)&=&\int_{0}^{2\pi} \int_{-1}^{1} {^{\small(-)}} Y_{l}^{m}(x,\phi) P_{(a)}(u,x,\phi) dx d\phi\\
&\equiv&{^{\small(-)}}\left<l,m|P_{(a)}\right>.
\end{eqnarray}}
Inserting the expanded solution (\ref{Pappr}) in the RT partial differential equation (\ref{Pdot}) we obtain
an autonomous dynamical system of dimension $(N_{P}+1)^2$ for the modal coefficients $(A_{l,m}(u),B_{lm}(u))$,
namely,
\begin{eqnarray}
\label{Glk}
\nonumber
{\dot{A}}_{l,m}(u)=\mathcal{A}_{l,m}\Big(A_{l,m}(u),B_{l,m}(u)\Big),\\
{\dot{B}}_{l,m}(u)=\mathcal{B}_{l,m}\Big( A_{l,m}(u),B_{l,m}(u)\Big),
\end{eqnarray}
where ${\mathcal{A}}_{l,m}$ and ${\mathcal{B}}_{l,m}$ are polynomials of order $(N_{P}+1)^2$ in $A_{l,m}(u)$ and $B_{l,m}(u)$
and a dot denotes here $u$-derivative. The Galerkin scheme guarantees that the projections of $(P(u,x,\phi)-P_{(a)}(u,x,\phi))$
onto each basis function, namely, $<(P(u,x,\phi)-P_{(a)}(u,x,\phi)),Y_{l}^{m}(x,\phi)>$ approach zero when $N_P \rightarrow \infty$
so that (\ref{Pappr}) approaches the exact solution $P(u,x,\phi)$, in the sense of the norm of the projection basis space of the
SHs.
%
The initial conditions $(A_{l,m}(u_0),B_{lm}(u_0))$ to be used to integrate the dynamical system (\ref{Glk})
are provided by the initial data $K(u_0,x,\phi)$ constructed in last Section (Eqs. (\ref{eq3_13-i})).
These initial values are obtained from the Galerkin decomposition of $P(u_0,x,\phi)$ (cf. Eq. (\ref{Pappr}))
evaluated according to (\ref{modal}).
\par As a matter of fact, the projections that transform the RT equation into the dynamical
system (\ref{Glk}) demands actually that we replace $P(u,x,\phi)$ by $P_{(a)}(u,x,\phi)$ in (\ref{lambdaP}) and (\ref{Pdot}).
But in doing so we excessively increase the computational demand, overloading the computer data storage. To circumvent this problem,
we divide the RT equation into two parts that we called constraints: the $\lambda$-constraint and the
$P^3$-constraint. The process will consist then in also expanding -- in the same way as in (\ref{Pappr}) -- the expressions for $\lambda$
and $P^3$,
{\small
\begin{eqnarray}
\label{lambdaappr}
\nonumber
\lambda_{(a)}(u,x,\phi)=\sum_{l=0}^{N_{\lambda}} \frac{1}{2} E_{l,0}(u) {^{\small(+)}} Y_{l}^{0}(x,\phi)+\\
\nonumber
+\sum_{l=1}^{N_{\lambda}} \sum_{m=1}^{l} E_{l,m}(u){^{\small(+)}} Y_{l}^{m}(x,\phi)+\\
+\sum_{l=1}^{N_{\lambda}} \sum_{m=1}^{l} F_{l,m}(u){^{\small(-)}} Y_{l}^{m}(x,\phi),
\end{eqnarray}}
{\small
\begin{eqnarray}
\label{P3appr}
\nonumber
{P^3}_{(a)}(u,x,\phi)=\sum_{l=0}^{N_{P^3}} \frac{1}{2} C_{l,0}(u) {^{\small(+)}} Y_{l}^{0}(x,\phi)+\\
\nonumber
+\sum_{l=1}^{N_{P^3}} \sum_{m=1}^{l} C_{l,m}(u){^{\small(+)}} Y_{l}^{m}(x,\phi)+\\
+\sum_{l=1}^{N_{P^3}} \sum_{m=1}^{l} D_{l,m}(u){^{\small(-)}} Y_{l}^{m}(x,\phi).
\end{eqnarray}}
\noindent Here, $N_{\lambda}$ and $N_{P^3}$ are integers that define the truncation orders for the respective expansions.
Throughout the paper, we take $N_{P}=N_{\lambda}=N_{P^3}$, but these orders are totally independent and can
be taken with diferent values. Now, we have a new two set of modal coefficients ($\{C_{l,m},D_{l,m}\}$ and
$\{E_{l,m},F_{l,m}\}$) that will be uniquely determined by the first employed set of modal coefficients
$\{A_{l,m},B_{l,m}\}$ by
{\small
\begin{eqnarray}
\label{newmodal}
\nonumber
C_{l,m}(u)&\equiv&{^{\small(+)}}\left<l,m|{P^3}_{(a)}\right>\\
\nonumber
&=&C_{l,m}(\{A_{l,m}(u),B_{l,m}(u)\}),\\
\nonumber
D_{l,m}(u)&\equiv&{^{\small(-)}}\left<l,m|{P^3}_{(a)}\right>\\
\nonumber
&=&D_{l,m}(\{A_{l,m}(u),B_{l,m}(u)\}),\\
\nonumber
E_{l,m}(u)&\equiv&{^{\small(+)}}\left<l,m|{\lambda}_{(a)}\right>\\
\nonumber
&=&E_{l,m}(\{A_{l,m}(u),B_{l,m}(u)\}),\\
\nonumber
F_{l,m}(u)&\equiv&{^{\small(-)}}\left<l,m|{\lambda}_{(a)}\right>\\
&=&F_{l,m}(\{A_{l,m}(u),B_{l,m}(u)\}).
\end{eqnarray}}
The RT dynamical equation will have then a new form: the left side will be written
by the $u$-derivative of $P_{(a)}$
{\small
\begin{eqnarray}
\label{Pdotleft}
\nonumber
{\dot{P}}_{(a)}(u,x,\phi)_{left}=\sum_{l=0}^{N_{P}} \frac{1}{2} \dot{A}_{l,0}(u) {^{\small(+)}} Y_{l}^{0}(x,\phi)+\\
\nonumber
+\sum_{l=1}^{N_{P}} \sum_{m=1}^{l} \dot{A}_{l,m}(u){^{\small(+)}} Y_{l}^{m}(x,\phi)+\\
+\sum_{l=1}^{N_{P}} \sum_{m=1}^{l} \dot{B}_{l,m}(u){^{\small(-)}} Y_{l}^{m}(x,\phi),
\end{eqnarray}}
\noindent and, its right side, will be given by the replacement of $\lambda$ and $P^3$ in (\ref{Pdot})
by the constraint expressions (\ref{lambdaappr}) and (\ref{P3appr}).
{\small
\begin{eqnarray}
\label{Pdotright}
\nonumber
{\dot{P}}_{(a)}(u,x,\phi)_{right}=-\frac{{P^3_{(ap)}}}{12m_0}\Big[{\lambda_{(a)}}_{xx}-
2x{\lambda_{(a)}}_{x}+\frac{{\lambda_{(a)}}_{\phi\phi}}
{(1-x^2)}\Big].
\end{eqnarray}}

Finally we project both sides to get
the relation between the $u$-variation of the set $\{A_{l,m}(u),B_{l,m}(u)\}$ and the new modal coefficients
{\small
\begin{eqnarray}
\label{modaldot}
\nonumber
\dot{A}_{l,m}(u)&\equiv&{^{\small(+)}}\left<l,m|{\dot{P}}_{(a)}\right>\\
\nonumber
&=&\dot{A}_{l,m}(\{C_{l,m}(u),D_{l,m}(u),E_{l,m}(u),F_{l,m}(u)\})\\
\nonumber
\dot{B}_{l,m}(u)&\equiv&{^{\small(-)}}\left<l,m|{\dot{P}}_{(a)}\right>\\
&=&\dot{B}_{l,m}(\{C_{l,m}(u),D_{l,m}(u),E_{l,m}(u),F_{l,m}(u)\}).
\end{eqnarray}}
\noindent This completes our scheme to obtain the dynamical system version of the non-axisymmetric RT equation (\ref{Pdot}).
To solve it, we use a fourth-order Runge-Kutta recursive method adapted to our constraints. So we have to get
the initial values for the modal coefficients, say $\{A_{l,m}(u_0),B_{l,m}(u_0)\}$
{\small
\begin{eqnarray}
\label{initialmodal}
\nonumber
A_{l,m}(u_0)&\equiv&{^{\small(+)}}\left<l,m|{P_0}_{(a)}\right>,\\
B_{l,m}(u_0)&\equiv&{^{\small(-)}}\left<l,m|{P_0}_{(a)}\right>.
\end{eqnarray}}
The effect of the truncation $N_P$ on the initial data
may be evaluated by the relative error ${\cal{R}}(u_0)$ between the approximate and
exact expressions given by (\ref{Pappr}) and $P(u_0,x,phi)\equiv 1/K(u_0,x,phi)$ (cf. (\ref{eq3_13-i})),
\begin{eqnarray}
\label{eq13cc}
{\cal{R}}(u_0,x)= \frac{\mid P(u_0,x,\phi)-P_{(a)}(u_0,x,\phi) \mid}{P(u_0,x,\phi)}.
\end{eqnarray}
In all the numerical experiments of the paper we have adopted $N_{P}=7$;
with this truncation the relative error above is of the order of, or smaller than $10^{-8}$,
for all $-1 \leq x \leq 1$ and $0 \leq \phi \leq 2\pi$.
The RT dynamics furnishes us with a further test to check the accuracy and reliability of our numerical codes.
In fact, for any sufficiently smooth $K(u,x)$ we have that the quantity
{\small
\begin{eqnarray}
\label{eq12_00}
\zeta(u) =\int^{2 \pi}_{0} d \phi \int^{1}_{-1}P^{-2}(u,x,\phi)~dx,
\end{eqnarray}}
is conserved along the dynamics, namely, $\partial_{u} \zeta(u)=0$. Evaluating its exact value from
the initial data (\ref{eq3_13-i}) and at distinct steps of the computation we that $|\zeta(u_0)-\zeta(u)|\leq 10^{-10}$
for all our numerical experiments and for all the sampled values of $u>u_0$, for the adopted
truncation $N_{P}=7$.
\par Our numerical experiments are realized with the initial data (\ref{eq3_13-i}) -- which
corresponds to one initial black hole boosted along the $z$-axis and the second black hole moving along the direction
determined by ${\bf n}=(n_1,n_2,n_3)$ with $n_1^2+n_2^2+n_3^2=1$-- having in principle four independent parameters.
\par We vary $\alpha$ in the interval $(0,1.5)$ for $\gamma=0.2$ and several values of $n_1$.
Exhaustive numerical experiments show that for a sufficiently large computation time $u_f$ all modal coefficients become
constant, namely, at $u_{f}$ the emission of gravitational waves is considered to have ceased.
For $u_f$ we actually have that $|A_{l,m}(u_f+h)-A_{l,m}(u_f)|\leq 10^{-10}$, $|B_{l,m}(u_f+h)-B_{l,m}(u_f)|\leq 10^{-12}$ for all $l=0...7$,
where $h$ is the integration step.
\par From these modal coefficients $A_{l,m}(u_f)$ and $B_{l,m}(u_f)$ we reconstruct
the final configuration ${\small P(u_f,x,\phi) \simeq {P_{(a)}}(u_f,x,\phi)}$ that, in all cases, can be expressed as
{\small
\begin{eqnarray}
\label{finalconf}
\nonumber
P(u_f,x,\phi)& = &\frac{A_{0,0}(u_f)}{2}+\frac{A_{1,0}(u_f)}{2} \cos \theta\\
\nonumber
&+&A_{1,1}(u_f) \sin \theta \cos \phi\\
\nonumber
&+& B_{1,1}(u_f) \sin \theta \cos \phi +{\mathcal{O}}(10^{-10})\\
&\simeq &\frac{1}{K_{f}}(\cosh \gamma_{f}+ {\textbf{n}}_f \cdot {\hat \textbf{x}} ~\sinh \gamma_{f}).
\end{eqnarray}}
The rms error of the second equality in (\ref{finalconf}) is of the order of, or smaller that $10^{-12}$.
The expression in the second equality of (\ref{finalconf}) corresponds to a boosted black hole along the direction
${\bf n}_f=(n_{1f},n_{2f},n_{3f})$ with boost parameter $\gamma_f$ and rest mass  $m_0 K_{f}^{3}$, a confirmation of the
Chrusciel-Singleton theorem\cite{chrusciel}.
\par From (\ref{finalconf}) we can identify

{\small
\begin{eqnarray}
\label{finalconf1}
\nonumber
A_{0,0}(u_f)&=&\frac{2 \cosh \gamma_f}{K_f},\\
A_{1,0}(u_f)&=&\frac{2~ n_{3f}~ \sinh \gamma_f}{K_f},\\
\nonumber
A_{1,1}(u_f)&=&\frac{n_{1f}~\sinh \gamma_f}{K_{f}},\\
\nonumber
B_{1,1}(u_f)&=&\frac{n_{2f}~\sinh \gamma_f}{K_{f}},
\end{eqnarray}}
up to ${\mathcal{O}}(10^{-12})$, from which we can read the final parameters of the remnant stationary black hole $(K_f, \gamma_f,{\bf n}_{f})$.
It results
{\small
\begin{eqnarray}
\label{nf}
\nonumber
n_{1f}&=&\frac{2 A_{1,1}(u_f)}{A_{1,0}(u_f)} ~n_{3f},~~n_{2f}= \frac{2 B_{1,1}(u_f)}{A_{1,0}(u_f)} ~n_{3f}\\
n_{3f}&=& \Big(1+ (\frac{2 A_{1,1}(u_f)}{A_{1,0}(u_f)})^2+(\frac{2 B_{1,1}(u_f)}{A_{1,0}(u_f)})^2~\Big)^{-1/2}\\
\nonumber
\gamma_f&=&\tanh^{-1}(\frac{1}{n_{3f}}~ \frac{A_{1,0}}{A_{0,0}})\\
K_f&=&\frac{2}{A_{0,0}} \cosh \gamma_f.
\end{eqnarray}}
An alternative to evaluate the parameter $K_f$ is the use of the initial data in the conserved quantity
(\ref{eq12_00}), namely,
{\small
\begin{eqnarray}
\label{eq12_001}
K_{f}=\Big( \frac{1}{4 \pi}\int^{2 \pi}_{0} d \phi \int^{1}_{-1}K^{2}(u_0,x,\phi)~dx \Big)^{1/2}~~.
\end{eqnarray}}
The agreement is within 10 significant decimal digits. This is also a test of the accuracy of $(n_{1f})^{2}+(n_{2f})^{2}+(n_{3f})^{2}=1$.
\par One of the basic results to be extracted from our numerical experiments
are the values of ($K_{f},\gamma_{f},n_{3f}$), for each of the independent parameters ($\alpha,\gamma,{\bf n}$) of the
initial data. The former are the basic parameters of the remnant that -- together with the initial data of the system
and the function $K(u,\theta,\phi)$ for all $u_0\leq u \leq u_f$-- allow us to evaluate quantities which are
characteristic of the radiative transfer processes involved in the gravitational wave emission.
Finally we remark that the integration of the dynamical system (\ref{modaldot})) used a fourth-order Runge-Kutta recursive method
(adapted to our constraints) together with a C++ integrator. Unless otherwise stated, all our numerical results are restricted
to the choice $\gamma=0.2$, corresponding to the initial infalling collision velocity $v\simeq 0.1973$.

\section{The planar nature of the general non-head-on coliision}

From the numerical experiments done to check the long time evolution of the initial data (\ref{eq3_13-i})
and to obtain the basic parameters of the remnant black hole, we observe that the final direction
of remnant velocity is contained in the plane determined by the the two directions of the initial individual colliding black holes.
For illustration let us consider the evolution of the initial data (\ref{eq3_13-i}), with ${\bf n}_z=(0,0,1)$
corresponding to the direction of the initial colliding black hole (boosted along the positive $z$-axis), with ${\bf n}=(0.26, 0.07,0.963068)$
corresponding to the direction of the second initial colliding black hole and the several values of $\alpha=0.1,~0.1,~0.20,~0.3,~0.50$.
The resulting final direction of the remnant is given by (cf. (\ref{nf}))
{\small
\begin{eqnarray}
\label{planar01}
\nonumber
{\bf n}_f(\alpha=0.1) &\simeq& (1.363345 ~ 10^{-2},3.670544 ~ 10^{-3},9.999003~ 10^{-1}),\\
\nonumber
{\bf n}_f(\alpha=0.2)& \simeq& (3.064877~10^{-2},8.251592~10^{-3},9.994961~10^{-1}),\\
\nonumber
{\bf n}_f(\alpha=0.3)& \simeq& (0.000000~10^{-2},0.000000~10^{-3},9.000000~10^{-1}),\\
\nonumber
{\bf n}_f(\alpha=0.5) &\simeq& (1.197374~10^{-1},3.223699~10^{-2},9.922820~10^{-1}).
\end{eqnarray}}
In all cases we obtain
{\small
\begin{eqnarray}
\label{planar02}
{{\bf n}_z} \cdot {\bf n} \wedge {\bf n}_f \simeq 0.
\end{eqnarray}}
Within machine precision, (\ref{planar02}) is valid up to ${\mathcal{O}}(10^{-10})$, and it also holds
for all numerical experiments done in the paper considering a large domain of the initial data parameters,
implying that the dynamics is indeed planar, as discussed at the end of Section III. The result (\ref{planar02}) could also be considered
as an additional test for the accuracy of our numerical code.
\par This fundamental result -- which will be used to save computational efforts
in the dynamical system evolution of the RT non-axisymmetric dynamics -- has an important physical counterpart
connected to the fact that the linear momentum flux carried out by gravitational waves is confined to the plane of the initial collision.
In other words, the linear momentum of the merged system in the direction ${\bf{n}_z} \wedge {\bf{n}}$ is conserved,
as discussed in Section VIII.
\par Let us then consider a rigid rotation of the two initial black hole system about the direction of motion of the
first initial black, namely, the $z$-axis. This rotation will be chosen so that the plane of the initial collision,
determined by the two directions ${\bf n}_z=(0,0,1)$ and ${\bf n}$, coincide with the $x-z$ plane. In this case
the direction of motion of the second initial colliding black hole is now expressed by the unit vector
${\bf{n}}=(n_1,0,n_2)$, with $n_1^2+n_3^2=1$. In this instance the initial data (\ref{eq3_13-i}) can assume the simpler form
{\begin{widetext}
{\small
\begin{eqnarray}
K(u_0,\theta,\phi)=\Big(\frac{1}{\sqrt{{\cosh \gamma+ \cos \theta \sinh \gamma}}}+
\frac{\alpha}{\sqrt{{\cosh \gamma - (\cos \rho_0 ~\cos \theta+ \sin \rho_0~ \sin\theta \cos \phi)\sinh \gamma}}}\Big)^2~.
\label{eq3_13-ii}
\end{eqnarray}}
\end{widetext}}
This initial data contains now three independent parameters $(\alpha, \gamma, \rho_0)$ only,
where $\rho_0$ is the angle formed by ${\bf{n}}$ and the $z$-axis and is used to parametrize
$n_1= \cos \rho_0$ and $n_2= \sin \rho_0$. Again, as already established, the motion of the resulting remnant
black hole will then be restricted to the $x-z$ plane for any values of the initial data parameters.
From the preceding discussions of the present Section we have that the initial data (\ref{eq3_13-ii}) describes
a general non head-on collision of two black holes in the realm of RT dynamics.
\par Now, for exhaustive numerical tests with the initial data (\ref{eq3_13-ii})
and a large range of parameters $(\alpha,\rho_0)$ we verify that: (i) the initial modal coefficients $B_{l,m}(u_0) \lesssim 10^{-17}$,
and (ii) its evolution via the RT dynamical system (\ref{modaldot}) maintains $B_{l,m}(u) \lesssim 10^{-17}$, for all $(l,m)$ in their allowed
range, and for all $u_0 <u \leq u_f$. This corresponds -- within the precision of our computation -- to $B_{l,m}(u) =0$ for all u.
Obviously, from the second equation (\ref{nf}), we have that the unit vector determining the direction of the remnant will
always have the form ${\bf n}=(n_{1f},0,n_{3f})$.
Therefore our computational task can be simplified by taking all the $B_{l,m}$ coefficients equal to zero, which is equivalent
to restrict all our expansions in the Galerkin method decomposition to the cosine series only.
In the remaining of the paper we have adopted this procedure
in the numerical evaluations of RT dynamics for the data (\ref{eq3_13-ii}).
We have however made sample tests by verifying the absolute differences in the physical results originating from data either
from the complete decomposition or the decomposition using the cosine series only. The difference remains always of the order of,
or smaller than $10^{-17}$. The angle $\rho_0$ ($0\leq \rho_0 \leq \pi/2$) will be
denoted the collision angle, the limiting cases $\rho_0=0$ corresponding to a head-on collision and
$\rho_0=\pi/2$ corresponding to a right angle collision.

\par It will be possible to follow the full evolution of (\ref{eq3_13-ii}) by considering from very small $\alpha$
up to those values for which the nonlinearities start to play an important role in the dynamics. We will restrict
ourselves to the range $0 < \alpha \leq 1.2$ The range of the initial incidence angle in the numerical experiments
will be $0 \leq \rho_0 \leq 90^{o}$, the limits corresponding respectively to a head-on collision and an orthogonal
collision.
\section{The Bondi-Sachs Four Momentum and Conservation Laws}
\par RT spacetimes describe the asymptotically flat exterior gravitational field of a bounded system
radiating gravitational waves and in this sense they are in the realm of the 2+2 Bondi-Sachs formulation
of gravitational waves in General Relativity\cite{bondi, sachs}. Furthermore initial data for RT dynamics
are prescribed on null characteristic surfaces. Therefore suitable expressions for the physical quantities
to be used in the description of gravitational wave emission processes and its conservation laws must be derived.
To exhibit such expressions it is necessary to perform a coordinate transformation
from RT coordinates used in (\ref{eq1}) to a coordinate system in which the metric coefficients
satisfy the Bondi-Sachs boundary conditions. It should be noticed that in the RT coordinate system
the presence of the term $2 r {{K}_u}/K$ does not fulfill the appropriate boundary conditions.
Although the coordinate transformations from RT coordinates to Bondi-Sachs coordinates
cannot be expressible in a closed form (they are given by an infinite series in powers of $r^{-1}$)
\cite{unti}, their asymptotic expansion allows us to obtain the form of the required physical quantities.
In this section we will restrict ourselves to the Bondi-Sachs energy-momentum for the RT spacetimes
as well as its conservation laws. Our derivation for the non-axisymmetric case\cite{aranha4} follows
closely the work of G\"ona and Kramer\cite{kramer} done for the axisymmetric case.
\par From the supplementary vacuum Einstein equations $R_{UU}=0$, $R_{U \Theta=0}$, and $R_{U \Phi=0}$ in the
$2+2$ Bondi-Sachs formulation\cite{bondi,sachs} (where $(U,R,\Theta,\Phi)$ are the Bondi-Sachs coordinates),
we obtain in RT coordinates
{\small
\begin{eqnarray}
\label{eq29}
\nonumber
&&\frac{\partial m(u,\theta,\phi)}{\partial u}= - K \Big( {c_u^{(1)}}^2 + {c_u^{(2)}}^2 \Big)
+ \frac{1}{2}\frac{\partial}{\partial u}
\Big[3 c_{\theta}^{(1)} \cot \theta \\&+& 4 c_{\phi}^{(2)} \frac{\cos \theta}{\sin^2 \theta} -2 c^{(1)} + c_{\theta \theta}^{(1)}+ \frac{2}{\sin \theta}
 c_{\theta \phi}^{(2)} - \frac{1}{\sin^2 \theta} c_{\phi \phi}^{(1)} \Big]
\end{eqnarray}}
where $m(u,\theta,\phi)$ is the Bondi mass function and $c_u^{(1)}(u,\theta,\phi)$ and $c_u^{(2)}(u,\theta,\phi)$ are the two {\it news}
functions for the non-axisymmetric case\cite{aranha4}, corresponding to the two modes of polarization of the gravitational waves.
The extra factor $K$ in the first term of the second-hand-side of Eq. (\ref{eq29}) comes from the transformation
{\small
\begin{eqnarray}
{\rm lim}_{r \rightarrow \infty}\frac{\partial U}{\partial u}= \frac{1}{K},
\label{eq30}
\end{eqnarray}}
$U$ being the Bondi time coordinate. For the {\it news} satisfying the appropriate boundary conditions
$c^{(1)}=c^{(2)}=0$ and $c^{(1)}_{\theta}=c^{(2)}_{\theta}=0$ at $\theta=0, \pi$~,
we obtain the Bondi-Sachs four-momentum conservation
{\small
\begin{eqnarray}
\label{eq31}
\frac{d P^{\mu}(u)}{d u}= P_{W}^{\mu},
\end{eqnarray}}
where the Bondi-Sachs four-momentum $P^{\mu}(u)$ is defined as
{\small
\begin{eqnarray}
\label{eq32}
P^{\mu}(u)= \frac{1}{4 \pi} \int^{2 \pi}_{0} d \phi \int^{\pi}_{0} m(u,\theta,\phi)~l^{\mu} \sin \theta~ d \theta,
\end{eqnarray}}
and
{\small
\begin{eqnarray}
\label{eq33}
P_{W}^{\mu}(u)= -\frac{1}{4 \pi} \int^{2 \pi}_{0} d \phi \int^{\pi}_{0} K~l^{\mu} \Big( {c_u^{(1)}}^2 + {c_u^{(2)}}^2 \Big) \sin \theta~ d \theta,~~
\end{eqnarray}}
corresponds to the net flux of energy-momentum carried out by the gravitational waves emitted.
In the above $l^{\mu}=(1,-\sin \theta \cos \phi, -\sin\theta \sin\phi,-\cos\theta)$ is a null vector
relative to an asymptotic Lorentz frame at infinity. We note that the last
term in (\ref{eq29}) vanishes in the integrations due to the boundary conditions of the news.
\par For $\mu=0$, Eq. (\ref{eq31}) yields the Bondi mass formula
{\small
\begin{eqnarray}
\label{eq33-I}
\frac{d M_B(u)}{du}=-P_W(u)
\end{eqnarray}}
where $M_B(u)$ is the Bondi mass at a time $u$ and
{\small
\begin{eqnarray}
\label{eq33v}
P_W(u)=\frac{1}{4 \pi}  \int^{2 \pi}_{0} d \phi \int^{\pi}_{0} K  \Big( {c_u^{(1)}}^2 + {c_u^{(2)}}^2 \Big) \sin \theta d \theta.~~~
\end{eqnarray}}
is the power extracted from the system by the gravitational wave emission in a time $u$. The total energy $E_W$ carried
out of the system by the gravitational waves is given by the time integral of (\ref{eq33v}) up to $u_f$, and corresponds to the
total Bondi mass extracted from the system, $M_B(u_0)-M_B(u_f)=E_W$, where $M_B(u_f)=m_0 K^{3}_{f} \cosh \gamma_f$ is the Bondi mass of the remnant.

\par For $\mu=x,y,z$, Eqs. (\ref{eq31}) determine the liner momentum conservation of the system,
{\small
\begin{eqnarray}
\label{eq33vv}
\frac{d~{\bf P}(u)}{du} ={\bf P}_W(u)
\end{eqnarray}}
where the vector
{\small
\begin{eqnarray}
\label{eq33vvv}
{\bf P}_W(u)=\frac{1}{4 \pi}  \int^{2 \pi}_{0} d \phi \int^{\pi}_{0} K \hat{\bf x} \Big( {c_u^{(1)}}^2 + {c_u^{(2)}}^2 \Big) \sin \theta d \theta.~~
\end{eqnarray}}
is the net momentum flux carried out by the gravitational waves emitted.
In the above $\hat {\bf{x}}=(\sin \theta \cos \phi,\sin \theta \sin \phi ,\cos \theta )$.
\par We are now able to analyze the radiative processes that lead the merged system from its initial configuration
to the final configuration of the remnant black hole.

\section{The power extracted by the gravitational wave emission and the efficiency of the mass-energy radiative transfer: upper and lower bounds}
\begin{figure}[t]
\begin{center}
\hspace{-0.5cm}
\vspace{-0.5cm}
{\includegraphics*[height=7.8cm,width=9.5cm]{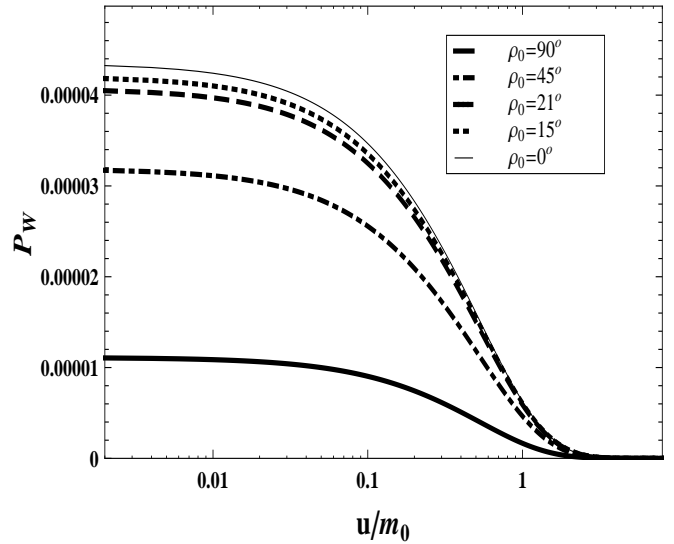}}
\caption{Linear-log plot of the energy flux (power) carried out by gravitational waves as a function of $u/m_0$,
for $\alpha=0.1$ and $\gamma=0.2$ and several values of the collision angle $\rho_0$. We see that, for
this relatively small value of $\alpha$, the gw emission corresponds to a pulse of short
duration, $\Delta u /m_0\sim 3.5$, and its initial intensity decreases as $\rho_0$ increases.
The case $\rho_0=0$ (head-on collision) constitutes an upper bound, while the orthogonal collision ($\rho_0=90^{o}$)
constitutes a lower bound for the total energy emitted (the area below the curve), in accordance with
the efficiency behavior.}
\label{figRad1}
\end{center}
\end{figure}
\begin{figure}[t]
\begin{center}
\hspace{-0.5cm}
\vspace{-0.5cm}
{\includegraphics*[height=7.8cm,width=9.5cm]{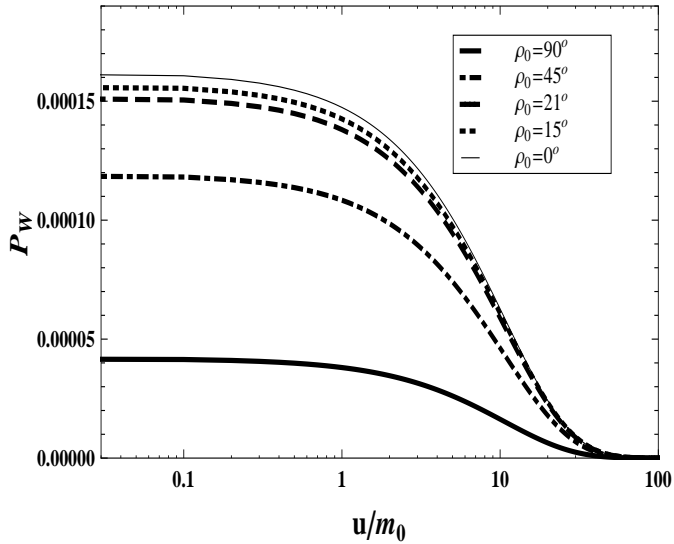}}
\caption{Linear-log plot of the energy flux (power) carried out by gravitational waves as a function of $u/m_0$,
for $\alpha=0.6$ and $\gamma=0.2$ and several vaues of the collision angle $\rho_0$. Analogous to the case of $\alpha=0.1$,
the head-on collision constitutes an upper bound, and the orthogonal collision a lower bound for the total energy emitted
in accordance with the efficiency behavior. The total energy emitted is spread on a larger interval of time, $\Delta u /m_0 \sim 72$.}
\label{figRad2}
\end{center}
\end{figure}

We now examine the energy extracted from the merged system by the gravitational waves. From our discussion in Section VI
on the Bondi-Sachs conservation laws, and specifically from Eqs. (\ref{eq33-I}) and (\ref{eq33v}) we have that the power emitted by the system
in a time $u$ is given by
{\small
\begin{eqnarray}
\label{power}
\nonumber
P_{W}(u) &\equiv& d E_W(u)/du\\
&=&\frac{1}{4 \pi} \int^{2 \pi}_{0} d \phi \int^{\pi}_{0} K  \Big( {c_u^{(1)}}^2 + {c_u^{(2)}}^2 \Big) \sin \theta d \theta.~
\end{eqnarray}}
In Figs. \ref{figRad1} and \ref{figRad2} we plot the power emitted $P_W(u)$ as a function of
$u/m_0$ for $\alpha=0.1$ and $\alpha=0.6$, respectively, and $\gamma=0.2$ fixed, and for several values of the incidence angle $\rho_0$.
The total energy emission $E_W(u_f)$ in each case is measured by the area below the respective curve -- in accordance with (\ref{eq33v}).
Two important features are to be noted in the Figures. First, we can see that a head-on collision ($\rho_0=0$) constitutes
an upper-bound for the total energy emitted, as well as the orthogonal collision ($\rho_0=90^{o}$) constitutes a lower-bound
for this energy. This pattern is typical for any $0< \alpha <1 $ and for all $\gamma$.
Actually in our numerical experiments we verified
this important result for $0 \leq \alpha \leq 1$, consistent with the curves of the efficiency of mass-energy extraction by
gravitational waves, as we will discuss later in the present Section.
\par Second, although the curves present the same behavior for distinct $\alpha$'s,
we can see that for $\alpha$ small (cf. Fig. \ref{figRad1} for $\alpha=0.1$) the emission corresponds to a short pulse of gravitational waves.
We indeed obtain that, for any of the incident angles $\rho_0$ (here including the head-on collision), the initial
power emitted decreases by three orders of magnitude in an interval of time $\Delta u / m_0\sim 3.5$. On the contrary, for $\alpha=0.6$
(cf. Fig. \ref{figRad2}) the initial power emitted (which is one order of magnitude smaller than in the case $\alpha=0.1$) decreases
by three orders of magnitude in an much larger interval $\Delta u / m_0\sim 72$ for all the incidence angles $\rho_0$ considered.
In fact, analogous to the case of head-on collision analyzed in Ref. \cite{aranha1}, there is a threshold value of $\alpha \sim 0.67$
that separates regimes of short bursts of gravitational waves from a regime of quiescent long time emission.

\par For $\alpha$ in the interval $(0,1.5]$ and for several values of $\rho_0$ in the interval $(0,\pi/2)$
we have determined $(K_f,\gamma_f,\rho_{f})$ which characterize the final boosted black hole configuration.
From these parameters of the black hole remnant and the total radiated energy $E_W(u_f)$, the efficiency $\Delta$ of mass-energy
extraction by gravitational wave emission can be evaluated. According to Eardley\cite{eardley} the efficiency $\Delta$ is defined as
{\small
\begin{eqnarray}
\label{eq13_0}
\Delta \equiv \frac{\Big( M_{B}(0)-M_{B}(u_f) \Big )}{M_{B}(0)},
\end{eqnarray}}
where $M_{B}(u_0)$ is the initial Bondi mass and $M_{B}(u_f)$ is the final Bondi mass. We can
alternatively express $\Delta$ as
{\small
\begin{eqnarray}
\label{eq13_0}
\Delta =\frac{E_W(u_f)}{m_0 K_{f}^{3} \cosh \gamma_f+E_W(u_f)}.
\end{eqnarray}}
\par
\begin{figure}
\begin{center}
\hspace{-0.5cm}
{\includegraphics*[height=7.0cm,width=7.5cm]{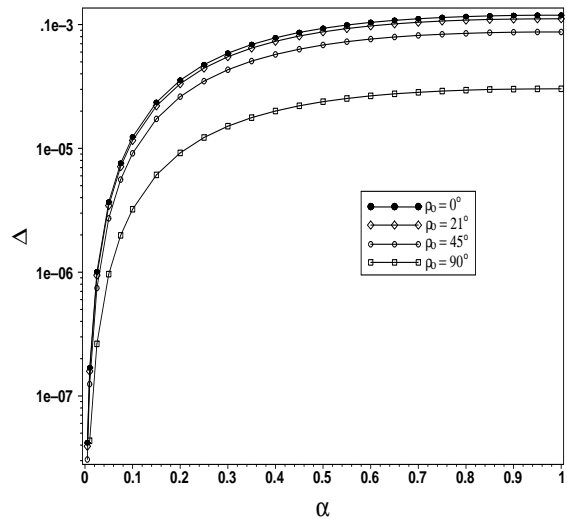}}
\vspace{0.2cm} \caption{Log-linear plot of the efficiency $\Delta$ for the range
$\alpha_1=(0,1]$ and several values of the incidence angle $\rho_0$. The initial parameter $\gamma=0.2$
is fixed but the distribution of points for other values of $\gamma$ exhibit an analogous behavior.
The points are connected for a better visualization. Values of $\Delta$ are redundant due to the relation $\Delta(\alpha)=\Delta(1/\alpha)$
for $\alpha \in (0, \infty)$.}
\label{fig1}
\end{center}
\end{figure}

\begin{figure}
\begin{center}
\hspace{-0.5cm}
{\includegraphics*[height=7.0cm,width=7.5cm]{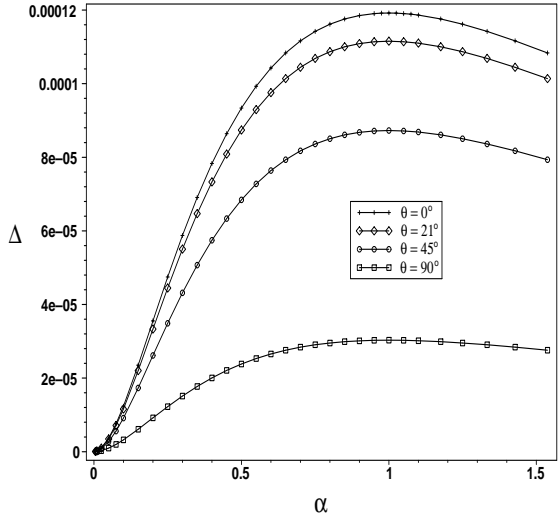}}
\vspace{0.2cm} \caption{Plot of the efficiency $\Delta$ in the range
$\alpha_1=(0,1.55)$ exhibiting the absolute maxima at $\alpha=1$ for $\rho_0=90^{o},~45^{o},~21^{o}$ and $0^{o}$.
The points are connected for a better visualizaton. This behavior of maxima at $\alpha=1$
actually holds for any $0 < \rho_0 \leq 90^{o}$ and any $\gamma$.}
\label{fig1-i}
\end{center}
\end{figure}

 In Fig.1 we display the log-linear plot of the efficiency $\Delta$  versus the mass ratio parameter $\alpha$
for $\gamma=0.2$ and several values of the collision angle $\rho_0$ varying between $\rho_0=0$ (head-on collision)
and $\rho_0=90^{o}$ (orthogonal collision). We see that the case of head-on collisions constitute an upper bound
for the efficiency $\Delta$ while the case of orthogonal collisions constitute a lower bound for $\Delta$,
in accordance with the power extraction behavior discussed above and illustrated in Figs. \ref{fig1} and \ref{fig1-i}.
The efficiencies also decrease as the initial angle of collison $\rho_0$ increases.
\par We can also derive that, due to the form of the initial data (\ref{eq3_13-ii}), the efficiency $\Delta$ has a maximum
for $\alpha=1$. In fact, let us consider $\alpha > 1$ in (\ref{eq3_13-ii}) and factorize $\alpha$ so that the initial data
will turn out to be the product of $\alpha^2$ times a new $K(u_0,\theta, \phi)$, the latter corresponding to the
initial data of two colliding black holes, one with mass parameter $(1/\alpha)$ boosted along the negative $z$ axis
and the other with mass parameters $1$ boosted along the direction ${\bf{n}}=(\sin \rho_0,0,\cos \rho_0)$. By a rigid counter
clockwise rotation of the $x-z$ plane, about the Cartesian axis $y$, of an angle $\pi-\rho_0$ the unity vectors $\bf{n}$
and ${\tilde{\bf{n}}}=(0,0,1)$ will transform respectively into ${\bf{n}}=(0,0,-1)$ and ${\tilde{\bf{n}}}=(-\sin \rho_0,0,-\cos \rho_0)$
reproducing (\ref{eq3_13-ii}) for $\alpha \rightarrow 1/\alpha$. Now it is not difficult to see that, under the rescale of
$K(u_0,\theta,\phi)$ by a factor $\alpha^2$, both $M_{B}(u_f)$ and $E_W(u_f)$ rescale with $\alpha^6$ so that the efficiency
$\Delta$ satisfies
{\small
\begin{eqnarray}
\label{Delta}
\Delta(\alpha)=\Delta(1/\alpha),
\end{eqnarray}}
for $\alpha \in (0,\infty)$.
The rescale of $E_W$ with $\alpha^6$ uses Eq. (\ref{eq33v}) and also that $d u$ rescales with $\alpha^8$, the latter derived directly from the RT equation.
\par Since we can see numerically that $\Delta$ increases with $\alpha$ in the range $\alpha=(0,1)$ it
necessarily follows that $\Delta$ has a maximum at $\alpha=1$ for any $0 < \rho_0 \leq 90^{o}$,
as illustrated in Fig. \ref{fig1-i}. This property holds also for any $\gamma$.
Also we have that the difference of the efficiencies between the limit configurations, $\rho_0=0^{o}$ (upper bound)
and $\rho_0=90^{o}$ (lower bound), increase as $\alpha$ increases, reaching a maximum at $\alpha=1$.
For the present case $\gamma=0.2$ we obtain numerically that $\Delta_{\rm max}(\rho_0=0^{o})\simeq 1.191990 \times 10^{-4}$,
$\Delta_{\rm max}(\rho_0=21^{o})\simeq 1.115352 \times 10^{-4}$,
$\Delta_{\rm max}(\rho_0=45^{o}) \simeq 0.872631 \times 10^{-4}$,
$\Delta_{\rm max}(\rho_0=90^{o}) \simeq 3.029724 \times 10^{-5}$.

\section{Gravitational wave recoil: the momentum flux carried out by gravitational waves and the momentum of the remnant}

The analysis of the linear momentum extraction from the merged system by the gravitational waves is made using the Bondi-Sachs conservation
laws (\ref{eq31}) for $\mu=x, y, z$, which reads
{\small
\begin{eqnarray}
\label{momentum}
{\bf P}_W(u)=\frac{1}{4 \pi}  \int^{2 \pi}_{0} d \phi \int^{\pi}_{0} K \hat{\bf x} \Big( {c_u^{(1)}}^2 + {c_u^{(2)}}^2 \Big) \sin \theta d \theta,~~
\end{eqnarray}}
where ${\bf P}_W(u)$ is the momentum flux carried out by the gravitational waves emitted. In accordance with the results and discussions of Section VI the linear momentum of the merged system orthogonal to the plane of collision is conserved, which in our choice of data (\ref{eq3_13-ii})
implies that $P_{W}^{y}=0$, namely, the net flux of linear momentum carried out by gravitational waves is restricted to the plane of collision.
Therefore from (\ref{momentum}) we have
{\small
\begin{eqnarray}
\label{momentumX}
{P}_{W}^{x}(u)&=&\frac{1}{4 \pi}  \int^{2 \pi}_{0}  d \phi \int^{\pi}_{0} \sin^2 \theta \cos \phi~ K \Big( {c_u^{(1)}}^2 + {c_u^{(2)}}^2 \Big) d \theta,~~~~~~\\
\label{momentumZ}
{P}_{W}^{z}(u)&=&\frac{1}{4 \pi}  \int^{2 \pi}_{0} d \phi \int^{\pi}_{0} \cos \theta \sin \theta~ K \Big( {c_u^{(1)}}^2 + {c_u^{(2)}}^2 \Big) d \theta,~~~~\\
\nonumber
\end{eqnarray}}
as the relevant momentum fluxes contributing in the conservation law (\ref{eq31}) and determining
the total impulse imparted on the merged system by gravitational wave emission.
\begin{figure}[t]
\begin{center}
\hspace{-0.2cm}
{\includegraphics*[height=5.50cm,width=7.8cm]{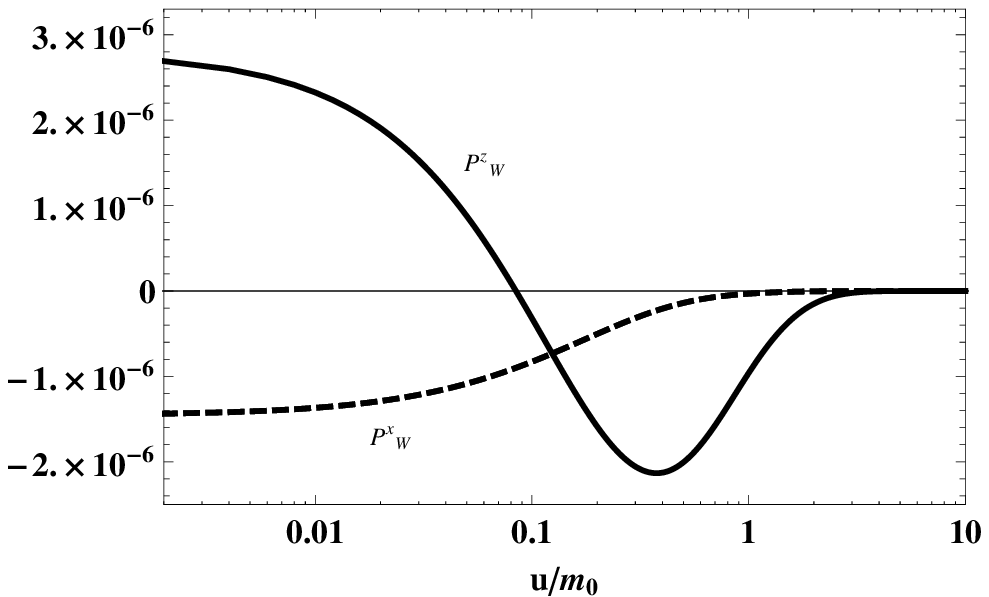}}
{\includegraphics*[height=5.50cm,width=7.8cm]{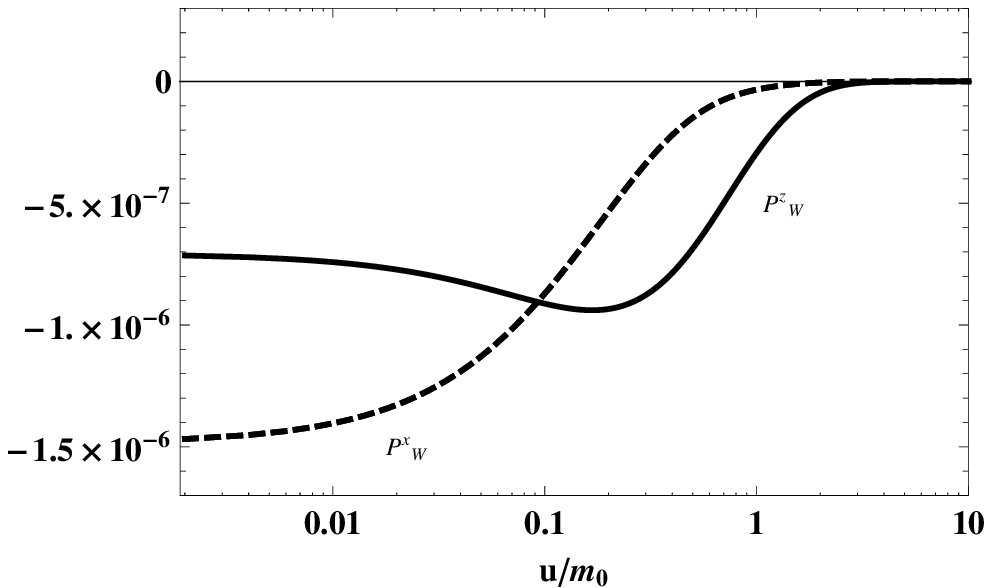}}
\vspace{0.1cm} \caption{Linear-log plots of the net momentum fluxes $P^{z}_{W}$ and $P^{x}_{W}$
carried out by the gravitational waves, for $\alpha=0.1$, $\gamma=0.2$, and incidence angles $\rho_0=15^{o}$ (top) and $\rho_0=90^{o}$ (bottom).
Due to the planar nature of the collision, the momentum flux $P_{W}^{y}$ is conserved.
For the case of the relatively small incidence angle $\rho_0=15^{o}$ (top) we have an initial regime with positive $P_{W}^{z}(u)$, $0< u < u_k$,
and a consequent increase of the linear momentum $z$-component of the merged system, $d P^{z}(u)/du >0$ (cf. (\ref{momentum})).
Along the $x$ axis the merged system is decelerated for all $u$ until the final configuration is reached.
For a large value of $\rho_0=90^{o}$ (bottom) the pattern changes and the system is decelerated along both directions $x$ and $z$.
The initial phase with $P_{W}^{z}(u)$ is not present for $\rho_0 \gtrsim 61^{o}$, as shown in Fig. \ref{patternII}.}
\label{patternI}
\end{center}
\end{figure}
\par
\begin{figure}[htb]
\begin{center}
\hspace{-0.2cm}
{\includegraphics*[height=5.50cm,width=7.8cm]{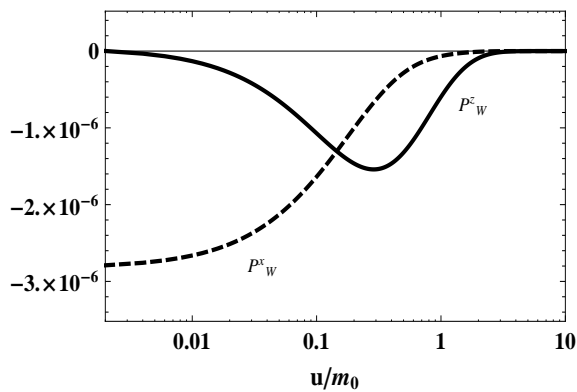}}
\vspace{0.1cm} \caption{Linear-log plots of the net momentum fluxes $P^{z}_{W}$ and $P^{x}_{W}$
carried out by the gravitational waves, for $\alpha=0.1$, and the initial incidence angle $\rho_0=61^{o}$ which corresponds
to a threshold configuration where the initial phase of positive net momentum flux along the $z$ axis not present, with $P_{W}^{z}(u_0)=0$.
In this case we can see that the total impulse imparted to the merged system along the $x$ direction is considerably
larger than the impulse in the $z$ direction.}
\label{patternII}
\end{center}
\end{figure}

 The total momentum conservation and the issue of kicks on the merged system were already analyzed in \cite{aranha2}
for the case of a head-on collision ($\rho_0=0^{o}$). For values of the incidence angle $\rho_0$ sufficiently small
the picture is analogous to that of a head-on collision, with $P_{W}^{x}(u) \simeq 0$ for all $u_0 < u < u_f$. However as $\rho_0$ increases
a negative net momentum flux appears along the $x$-direction leading to a decrease of the momentum of the merged system in that
direction for all $0 < u < u_f$. This is illustrated in Fig. \ref{patternI} (top) where the net momentum fluxes $P^{z}_{W}$ and $P^{x}_{W}$
carried out by the gravitational waves are plotted as a function of $u$, for mass ratio parameter $\alpha=0.1$ and incidence angle
$\rho_0=15^{o}$. We see that we have an initial phase where the momentum of the merged system along the $z$ axis increases, due to a
positive net momentum flux $P_{W}^{z}(u)$ for $u_0< u < u_k$, where $u_k / m_0 \simeq 0.085$ is the time when $P_{W}^{z}$ changes sign.
Therefore we have a dominant deceleration regime of the merged system until the remnant
black hole configuration is reached. For $\alpha=0.1$ fixed, this basic pattern is maintained up to $\rho_0 \simeq 61^{o}$, beyond
which both net gravitational wave momentum fluxes, $P_{W}^{x}(u)$ and $P_{W}^{z}(u)$, are negative for the whole domain of evolution
$u_0 \leq u \leq u_f$. Actually for $\rho_0 \gtrsim 61^{o}$ the initial positive net momentum flux along $z$ disappears, with the system
being decelerated along both directions $x$ and $z$ in the whole domain of the dynamical evolution. This is
illustrated in Fig. \ref{patternI} (bottom) for the limiting case $\rho_0=90^{o}$ (orthogonal collision).
The behavior of the threshold configuration $\rho_0 \simeq 61^{o}$ is illustrated in Fig. \ref{patternII}.
In a forthcoming paper we show that a basic role of the initial phase of positive net momentum flux $P_{W}^{z}(u)$
is to induce a large inspiral branch in the motion of the center-of-mass of the merged system.
\par The final momentum of the remnant can be obtained by evaluating (\ref{eq32}) at $u=u_f$. Taking into account that the
Bondi mass aspect for the final black hole remnant is given by
{\small
\begin{eqnarray}
\label{m-aspect}
m(u,\theta,\phi)=\frac{m_0 ~K_f^3}{(\cosh \gamma_f+({\bf n}_{f} \cdot {\hat {\bf x}} ) \sinh \gamma_f )^3},
\end{eqnarray}}
(cf. Eqs. (\ref{eq12}) and (\ref{finalconf})), we obtain by a straightforward integration that
{\small
\begin{eqnarray}
\label{eqPf}
P_{f}^{x}= n_{1f} ~ P_f,~~~~~ P_{f}^{z}= n_{3f}~ P_f,
\end{eqnarray}}
with modulus $P_f=m_0~K_{f}^{3} \sinh \gamma_f$. As expected we have $P_{f}^{y}$.
\par
\begin{figure}
\begin{center}
{\includegraphics*[height=6.5cm,width=8.8cm]{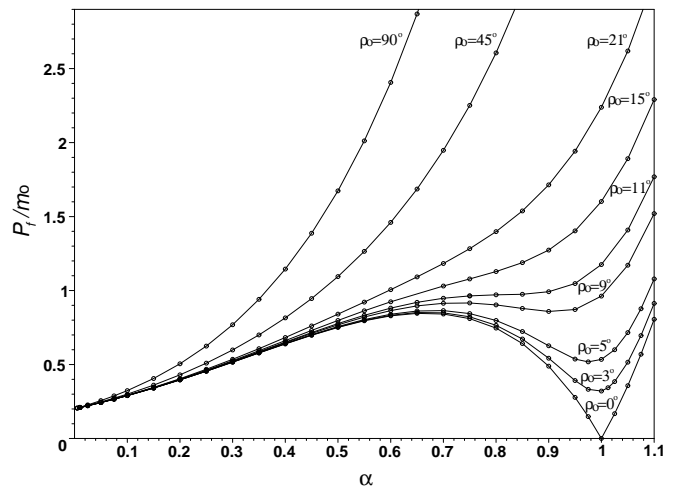}}
\vspace{0.2cm} \caption{Plot of the modulus $P_f$ of the momentum of the remnant black hole as a function of the mass ratio
parameter $\alpha$, for several values of the incidence angle of collision $\rho_0$. Points are connected for a better visualization.
Values of $P_f$ for $\alpha>1$ are redundant due to the relation $P_f(\alpha)= \alpha^6 P_f(1/\alpha)$.}
\label{fig2}
\end{center}
\end{figure}

 In Fig. \ref{fig2} we plot the distribution of the final momentum modulus $P_f$
of the remnant black hole as a function of $\alpha$, for several values of the incidence angle ranging
from $\rho_0=0$ (head-on collision) to $\rho_0=90^{o}$ (orthogonal collision).
We see that the head-on case ($\rho_0=0^{o}$) is a lower bound for the remnant moment distribution.
This is the only case in which the modulus of the final momentum of the remnant becomes zero, for $\alpha=1$,
as discussed in Ref. \cite{aranha2} where the axisymmetric dynamics was analyzed, making explicit the constrained character
of a head-on collision. For sufficiently small values of the incidence angle $\rho_0$ we see that the behavior follows that of
the head-on case, with the zero momentum at $\alpha=1$ being replaced by a minimum at $\alpha \lesssim 1$.
This minimum turns into a inflexion point for $\rho_0 \simeq 15^{o}$ and disappears for larger values of $\rho_0$.
Due to the dominant deceleration regime of the merged system, consequent of the linear momentum flux ${\bf P}_W$
extracted by the gravitational waves, the total momentum of the system decreases towards the distribution of $P_f$
given in Fig. \ref{fig2}. Finally, through an analogous analysis to the case of the efficiency $\Delta$, we have that
the final momentum $P_f$ satisfies the relation $P_f(\alpha)=\alpha^6 P_f(1/\alpha)$ for $\alpha \in (0,\infty)$.
Therefore in Fig. \ref{fig2} the values of $P_f$ for $\alpha >1$ are redundant.
\par
\begin{figure}
\begin{center}
\hspace{-0.2cm}
{\includegraphics*[height=6.0cm,width=8.5cm]{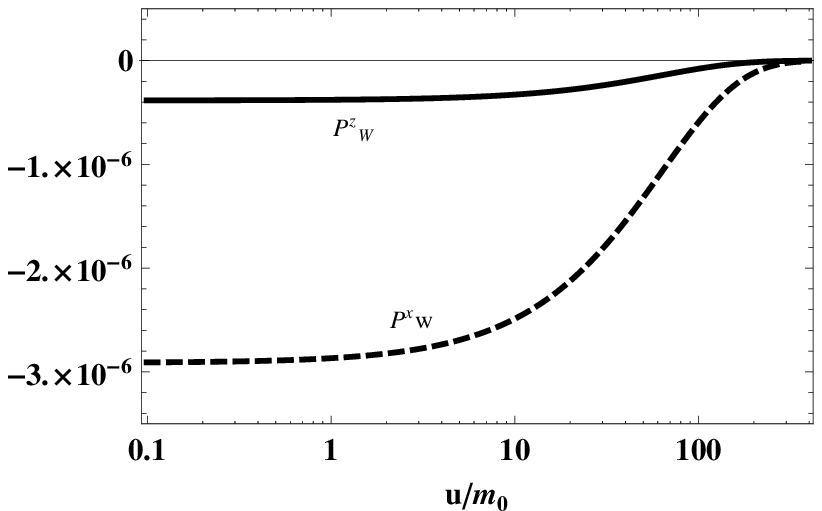}}
\vspace{0.1cm}
{\includegraphics*[height=6.0cm,width=8.5cm]{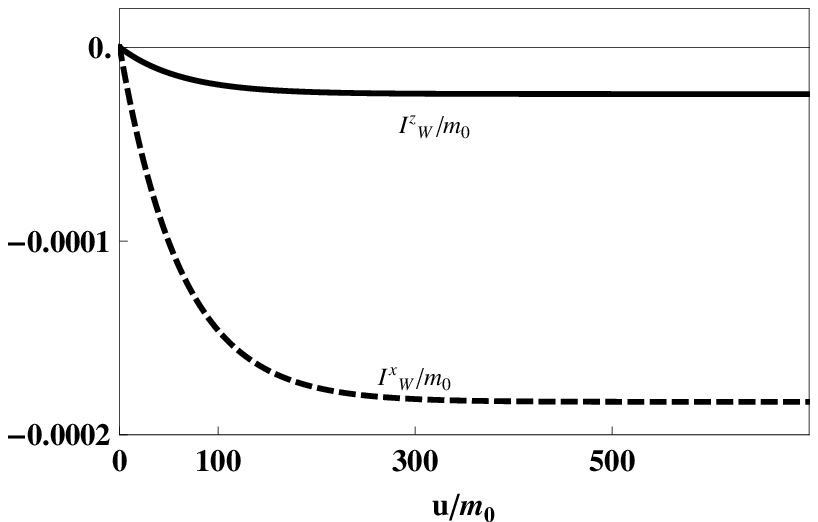}}
\vspace{0.1cm} \caption{Linear-log plot of the net momentum fluxes $P_{W}^{x}(u)$ and $P_{W}^{z}(u)$ (top), and the associated total
impulses $I_{W}^{x}(u)$ and $I_{W}^{z}(u)$ (bottom), for the case of equal-mass initial colliding black holes ($\alpha=1$) and
incidence angle $\rho_0=15^{o}$. The associated kick for this case is $V{k} \sim 0.9 {\rm km/s}$, this small value being
due to the value $\gamma=0.2$ adopted in our computation (cf. text).}
\label{ImpuseEqualMass}
\end{center}
\end{figure}

 A comment is in order now. An important new feature appears in non head-on collisions connected to the final momentum for equal
mass black holes ($\alpha=1$). Contrary to the case of a head-on collision and of the merging
of black hole binary inspirals, for a non head-on collision of two equal mass black holes the net gravitational
wave flux emitted is nonzero. This is the reason why the final momenta in Fig. \ref{fig2} for $\alpha=1$ is nonzero,
even if measured in an inertial frame with velocity ${\bf v}_{in}={\bf P}(u_0)/M_B(u_0)$ (namely, a zero-initial-Bondi-momentum frame)
relative to a rest inertial frame at infinity. This is illustrated in Figs. \ref{ImpuseEqualMass} ,
where we display the net momentum flux ${\bf P}_W(u)$
and the associated impulse ${\bf I}_W(u)$ for the case of equal-mass initial colliding black holes in a non head-on collision
with incidence angle $\rho_0=15^{o}$.
\par The analysis of the kick velocities generated in these processes of momentum extraction
(and the associated gravitational wave recoil) will be the subject of a future paper. In a related analysis
(cf. Ref. \cite{aranha2}) for head-on collisions we have shown that the distribution of kick velocities
as a function of the symmetric mass parameter fitted nicely Fitchett's law\cite{fitchett} whose original derivation was based
on post-Newtonian analytical estimates of gravitational wave emission in inspiral binaries. This corresponds to a Newtonian signature
in the dynamics and this result may suggest that even in the head-on post-merger phase, described by RT dynamics, a component of the
post-Newtonian dynamics of two interacting bodies emitting gravitational waves might be present/preserved. In the case of
non head-on collisions examined in the present paper a similar Newtonian signature appears involving the relation between the incident angle of
collision $\rho_0$ and the scattering angle of the remnant $\rho_f$, as we discuss in Section X.
\section{The angular wave pattern and the bremsstrahlung regime of the gravitational waves}

As discussed in Sec. II, the invariant characterization of a wave zone in RT spacetimes
and the consequent presence of gravitational waves is given by the functions $D(u,\theta,\phi)$ and $B(u,\theta,\phi)$,
Eqs. (\ref{eq8})-(\ref{eq9}), that characterize the curvature tensor components (\ref{eq7}) at the wave zone.
These functions correspond to the two modes of polarization of the gravitational waves and contain all the
information of the angular and time dependence of the gravitational wave amplitudes via the particular combination
{\small
\begin{eqnarray}
\label{penrose}
D(u,\theta,\phi)+i B(u,\theta,\phi) \sim ~(r \Psi_4),
\end{eqnarray}}
where $\Psi_4$ is the Weyl spinor associated with $N_{ABCD}/r$~\cite{penrose,aranhaTese}.
According to (\ref{eq8})-(\ref{eq9}), the quantity (\ref{penrose}) is specified once we
have the function $K(u,\theta,\phi) \equiv 1/P(u,\theta,\phi)$, which is in turn numerically obtained via the
approximation (\ref{Pappr}).
\par
\begin{figure}[t]
\begin{center}
{\includegraphics*[height=7.5cm,width=7.5cm]{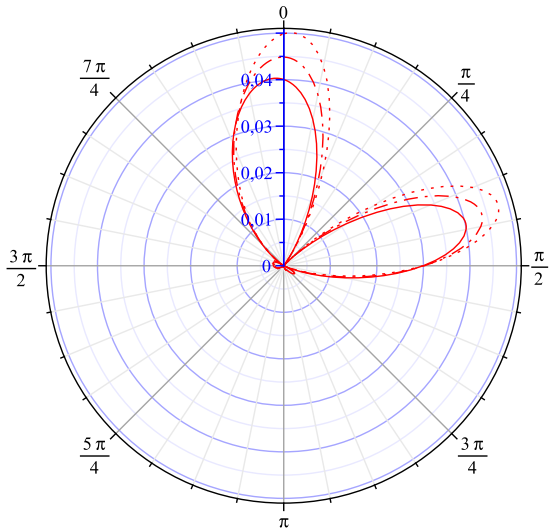}}
\caption{Polar plot of $\sqrt{D^2+B^2}$ (section by the plane $\phi=0^{o}$, corresponding to the plane of the collision) for times $u=0.01$ (dotted),
$u=0.05$ (dash-dotted) and $u=0.1$ (continuous), and initial data parameters
$\alpha=0.2$, $\rho_0=55^{o}$ and $\gamma=0.5$. The Figure shows a typical bremsstrahlung pattern, corresponding to a
strong deceleration regime at early times, with two dominant lobes
along the direction of motion of the merged system. The cone enveloping the two dominant lobes opens up as $u$ increases.
For larger values of $\gamma$ the enveloping cone of the forward lobes becomes more sharp.
The $z$ direction corresponds to the vertical axis.}
\label{Plot1}
\end{center}
\end{figure}

 In Fig. \ref{Plot1} we display the polar plots of $\sqrt{D^2+B^2}$ at early times $u=0.01$ ({\rm dotted}),$u=0.05$ ({\rm dash-dotted}) and $u=0.1$ ({\rm continuous}), and initial data parameters
$\alpha=0.2$, $\rho_0=55^{o}$ and $\gamma=0.5$, with section by the plane $\phi=0^{o}$ (corresponding to the plane of collision $x-z$).
The plots in Fig. \ref{Plot1} show, for each time, a pattern with two dominant lobes in the forward direction of motion of the merged system,
the first quadrant of the $x-z$ plane. The direction of the Bondi momentum vector at $u=0.01$ makes an angle $\Theta_B \simeq 8,43^{o}$
with the $z$ axis. The pattern is typical of a bremsstrahlung process due to the deceleration of the merged system,
analogous to the electromagnetic bremsstrahlung of a charge decelerated along its direction of motion.
As time increases we observe that the cone enveloping the dominant lobes opens up and the amplitudes decrease. For later
times (cf.Fig. \ref{Plot3} for $u=5.0$) the pattern evolves to the expected quadrupole structure with a much smaller amplitude.
We mention that the increase of the initial boost parameter $\gamma$ would sharpen the forward cone enveloping
of the two dominant lobes in the early regime, as expected in a ultrarelativistic configuration.
In our computations we fixed $m_0=10$. In the Figures the $z$ direction corresponds to the vertical axis.
vertical axis).
\begin{figure}
\begin{center}
{\includegraphics*[height=7.5cm,width=7.5cm]{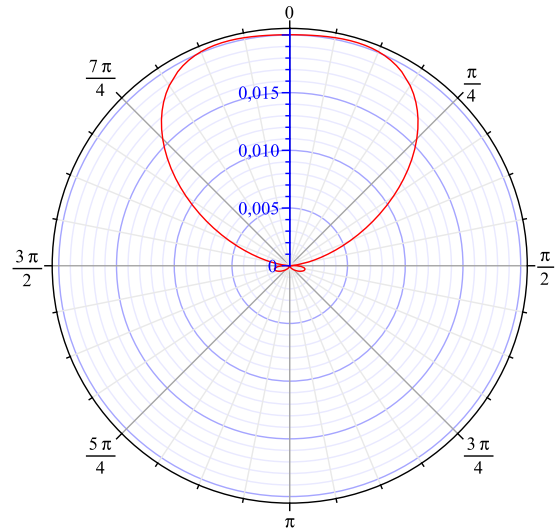}}
\caption{Polar plot of $\sqrt{D^2+B^2}$ (section by the plane $\phi=90^{o}$, corresponding to the plane $y-z$,
orthogonal to the plane of collision) for a time $u=0.1$, and initial data parameters
$\alpha=0.2$, $\rho_0=55^{o}$ and $\gamma=0.5$.
The symmetry about the $z$-axis is in accordance with the conservation of $P_{W}^{y}(u)=0$. Although gravitational waves
are emitted outside the plane of collision, the zero net momentum flux of this radiation component is consistent with the planar nature of the collision.}
\label{Plot2}
\end{center}
\end{figure}
\begin{figure}
\begin{center}
{\includegraphics*[height=7.5cm,width=7.5cm]{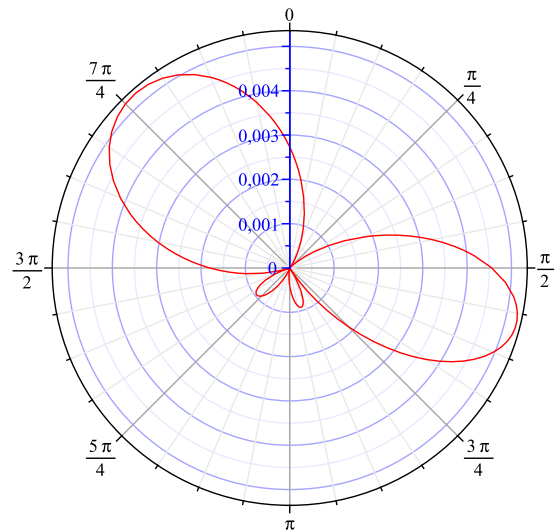}}
\caption{Polar plot of $\sqrt{D^2+B^2}$ (section by the plane $\phi=0^{o}$) for a time $u=5.0$, and same initial data parameters
of the previous Figures, showing the opening of the lobes and the setting already of the final quadrupole angular pattern.}
\label{Plot3}
\end{center}
\end{figure}
In Fig. \ref{Plot2} we show the polar plot of $\sqrt{D^2+B^2}$, with section by the plane
$\phi=90^{o}$ (corresponding to the plane $y-z$, orthogonal to the plane of collision), at $u=0.1$. The same initial
data parameters of the previous Figure were used. As expected the pattern is symmetric about the $z$ axis
in accordance with the conservation of $P_{W}^{y}(u)=0$. We then see that, although gravitational waves are emitted
outside the plane of the collision, this radiation component has a zero net momentum flux. Therefore it does not extract momentum of the
system, consistent with the planar nature of the collision.
\par Finally in Fig. \ref{Plot3} we display the polar plot of $\sqrt{D^2+B^2}$ (section by the plane $\phi=90^{o}$)
for a time $u=0.5$ and for the same initial data parameters of Fig. \ref{Plot1}. We can see the opening of the dominant lobes
and the forming of the final quadrupole angular pattern.

\section{A Newtonian signature in the relation between the incidence angle and the scattering angle
of the remnant}

One of the important parameters of the remnant black hole obtained from our numerical experiments is
the angle $\rho_f$ -- defined as $\rho_f= \cos^{-1} (n_{3f})$ and denoted the scattering angle of
the remnant -- resulting from the evolution of initial data with initial incidence angle $\rho_0$ .
In general, for the mass ratio parameter $\alpha$ sufficiently small, $\rho_f$ is much smaller
than the incidence angle $\rho_0$. However for large values of $\alpha$ this behavior changes
drastically as is the case for equal or nearly equal mass black holes.
%
\par
\begin{figure}
\begin{center}
\hspace{-0.5cm}
{\includegraphics*[height=4.0cm,width=7.5cm]{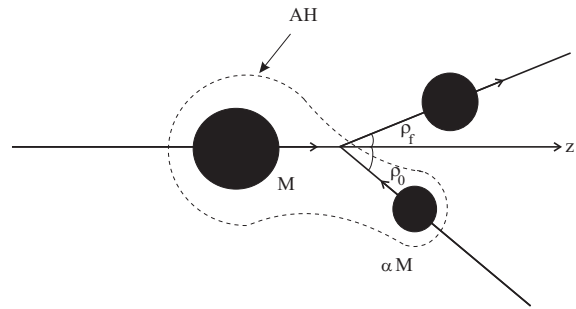}}
\vspace{0.2cm} \caption{Schematic diagram of collision of two boosted black holes with
initial infalling velocity parameter $\gamma$ and mass ratio $\alpha$. The dashed line depicts the
common apparent horizon enclosing the two colliding black holes.}
\label{angles}
\end{center}
\end{figure}

 No simple analytical relation between $\rho_0$ and $\rho_f$ was possible to be obtained from
the Bondi-Sachs conservation laws. However, as we will show,
the scattering angle can be related to the incidence angle by a formula that closely
approximates the Newtonian relation of angles in the
nonelastic scattering of classical particles, and which depends basically only on the mass ratio $\alpha$.
To see this let us consider the schematic diagram of collision shown in Fig. \ref{angles}. There
we have the two initial colliding black holes, one with total mass $M$ boosted along the positive $z$ axis
and the other with total mass $\alpha M$, with $\alpha < 1$, bosted along a direction making an
angle $\rho_0$ with the $z$ axis, in accordance with the asymptotic result (\ref{eq3_12}).
In this picture we consider the approximation in which we neglect the contribution of the initial
gravitational wave content to the individual masses, although
this contribution is small and in any case can be isolated. The direction of the momentum of
the remnant black hole makes an angle $\rho_f$ (the scattering angle) with the $z$ axis. Furthermore,
although in RT dynamics the presence a global apparent horizon is already present at
the initial time and the system may be viewed as a deformed black hole, we are also assuming that inside the
event horizon we still have two individual colliding black holes.
\par Under the above assumptions and taking into account that the dynamics of the collision involves
mass and momentum loss due to the emission of gravitational waves, the balance of linear momentum
can then be approximately expressed as
{\small
\begin{eqnarray}
\label{scat1}
\nonumber
M \sinh \gamma ~(1-\alpha \cos \rho_0)= (\delta_z)~ m_0 K_{f}^{3} \cos \rho_f \sinh \gamma_f,\\
\nonumber
M \sinh \gamma~ (\alpha \sin \rho_0)= (\delta_x)~ m_0 K_{f}^{3} \sin \rho_f \sinh \gamma_f,
\end{eqnarray}}
where $\delta_z$ and $\delta_x$ are correction factors introduced to account for the mass and
momentum loss in the process. The ratio of the above equations yield the relation
{\small
\begin{eqnarray}
\label{scat2}
\frac{\alpha \sin \rho_0}{1- \alpha \cos \rho_0}= \delta ~ \tan \rho_f
\end{eqnarray}}
where $\delta \equiv (\delta_x/\delta_z)$. Eq. (\ref{scat2}) provides a relation between the initial incidence angle
and the scattering angle of the remnant and must be validated by the results of the numerical experiments
with the initial data (\ref{eq3_13-ii}). We note that (\ref{scat2}) reproduces the equation for the inelastic
collision of classical particles when $\delta=1$.
\par Numerical results for $\rho_f$ obtained from initial data sampled in the whole range $0 < \alpha \leq 1$
shows that the parameter $\delta \sim 1$, its discrepancy from the
Newtonian classical value $1$ occurring for relatively small values of $\alpha$, where the asymmetry in the data is large.
For illustration
{\small
\begin{eqnarray}
\label{scat2-i}
\nonumber
(\alpha&=&0.1,~~\rho_0=45^{o},~~\rho_f \simeq 4.16^{o},~~ \delta \simeq 1.046047)\\
\nonumber
(\alpha&=&0.6,~~\rho_0=45^{o},~~\rho_f \simeq 35.78^{o},~~ \delta \simeq 1.022573)\\
\nonumber
(\alpha&=&1.0,~~\rho_0=45^{o},~~\rho_f \simeq 67.50^{o},~~ \delta \simeq 1.000000)
\end{eqnarray}}
As $\alpha$ increases $\delta \rightarrow 1$, the limiting
value $\delta=1$ being attained for $\alpha=1.0$ and any $\rho_0$,
yielding consequently (for $\alpha=1$)
{\small
\begin{eqnarray}
\label{scat3}
\rho_f=\frac{(180^{o}-\rho_0)}{2},
\end{eqnarray}}
with error of the order of, or smaller than $10^{-8}$, for all experiments made with $\alpha=1$.
In fact, $\rho_f=(180^{o}-\rho_0)/2$ is an exact solution for the Newtonian inelastic collision equation
(namely, Eq. (\ref{scat2}) with $\delta=1$). It is also worth remarking that  for equal-mass initial colliding black holes
($\alpha=1$) the direction determined by the angle $(180^{o}-\rho_0)/2$ with the $z$ axis corresponds to
an axis of symmetry of the initial data (cf. Fig. \ref{angles}).
\par Finally it can be of physical interest to compare the scattering angle with the initial momentum direction.
In our above approximation of the momentum conservation we obtain, by a straightforward calculation, that
the angle $\Theta_0$ of the initial momentum direction with the $z$-axis is given by
{\small $\tan \Theta_0 \equiv P^{x}(u_0)/P^{z}(u_0) = \alpha \sin \rho_0/(1- \alpha \cos \rho_0)$}, resulting from
the relation (\ref{scat2}) that {\small $\tan \Theta_0=\delta \tan \rho_f$}. This shows that the direction of motion
of the remnant presents a very small deviation from the direction of the initial momentum $\theta_0$,
as seen from an asymptotic inertial observer. This small difference is connected to the small values of the
impulse imparted to the merged system by the gravitational waves emitted. However these small values of the
impulse will be sufficient to generate kicks of a few hundred {\rm km/s} in the system.
\par The above results, relating $\rho_0$ and $\rho_f$, suggest a Newtonian signature present in the dynamics of the merged system with initial
data (\ref{eq3_13-ii}), indicating that the individuality of the two initial
colliding black holes is in some sense preserved in the initial deformed merged system.

\section{Conclusions and Final Remarks}

In this paper we have examined the numerical evolution of characteristic initial data
corresponding to non head-on collisions of two Schwarzschild black holes, in the realm
of Robinson-Trautman non-axisymmetric spacetimes. The Robinson-Trautman spacetimes present
already a global apparent horizon so that the dynamics covers a regime where the
merger has already set out. We have constructed initial data for the characteristic surface formalism
of Robinson-Trautman dynamics, that represent instantaneously this system. The RT equation is
integrated numerically using a Galerkin spectral method with a projection basis space in two
variables (the spherical harmonics). The planar nature of a generic non-head-on collision of two
black holes, which is restricted to the plane determined
by the two directions of motion of the two initial colliding black holes,
is used to test the accuracy of our numerical method and allows us to save a lot of computational effort.
The three basic independent parameters characterizing the initial data are the mass ratio parameter $\alpha$,
the boost parameter $\gamma$ (that defines the initial infalling velocity $v= \tanh \gamma$ of the initial individual
black holes) and the incidence collision angle $\rho_0$ of the two initial colliding black holes.
\par During the merger gravitational waves are emitted, extracting mass and linear
momentum of the system, until the remnant configuration is reached, when the gravitational wave emission ceases.
Typically the remnant is a boosted Schwarzschild black hole defined by its final rest mass $m_0 K_{f}^{3}$,
its final velocity parameter $v_f= \tanh \gamma_f$ and its direction of motion determined by the remnant
scattering angle $\rho_f$.
\par We evaluate the efficiency $\Delta$ of the mass-energy extraction as
a function of the mass ratio parameter $\alpha$ for several values of the incidence angle $\rho_0$.
We obtain that head-on collisions ($\rho_0=0^{o}$) constitute an upper bound,
while orthogonal collisions ($\rho_0=90^{o}$) constitute a lower bound for the efficiency.
This fact is connected to the behavior of the total gravitational wave energy emitted in the process
since the head-on collision and the orthogonal collision are configurations that correspond, respectively,
to an upper bound and a lower bound for the total energy emitted. The efficiency presents a maximum
at $\alpha=1$ for all values of the incidence angle, $0^{o} \leq \rho_0 \leq 90^{o}$, and for any boost parameter $\gamma$, which is
shown to be a consequence of some symmetries of the initial data (\ref{eq3_13-ii}).
\par The analysis of the momentum and energy flux of the gravitational waves is made using the Bondi-Sachs
energy-momentum conservation laws. The momentum extraction by gravitational waves with the consequent recoil of the system
is evaluated, with respect to an asymptotic Lorentz frame. For incidence angles $\rho_0$ up to a certain threshold value,
we have that the net momentum flux carried out by the gravitational waves along the $z$ axis is positive for a short initial phase
leading to an increase of the momentum of the merged system along this direction, $dP^{z}(u)/du >0$. This phase is followed
by a regime with negative net momentum flux that lasts until the final remnant configuration is attained.
For $\rho_{0}$ larger than the threshold value the initial phase of positive momentum flux is no longer present. For $\alpha=0.1$ the
threshold angle is $\rho_{0} \simeq 61^{o}$. Relative to the asymptotic Lorentz frame the Bondi-Sachs momentum components
$P^{z}(u)$ and $P^{x}(u)$ are positive for all $u$.
\par On the other hand, the net momentum flux along the $x$ direction is negative, for all times $u$ and incidence angle $\rho_0$,
so that the momentum of the system along the $x$ direction always decreases. We have that $P_{W}^{y}(u)=0$ for all $u$,
a consequence of the planar nature of the collision, so that the momentum of the merged system along this direction
is conserved (in fact we have $P^{y}(u)=0$ for all $u$).
\par In general we have a net strong deceleration of the system due to the emission of gravitational waves,
as a consequence of the the total impulse of the gravitational waves corresponding
to the integrated fluxes. Typically, for large $u \sim u_f$, the curves of the impulse tend to a negative
constant value (a plateau) associated with the final configuration when the gravitational wave emission ceases.
As a consequence, in accordance with the impulse conservation (\ref{eq33vv}) for $\Omega=u_f$,
there is a net momentum decrease due to the total impulse of the gravitational waves imparted
to the binary merged system. The importance of the initial acceleration of the
$z$-component of the momentum (when present) is to produce a large inspiral branch in the motion of the center-of-mass
of the system, as seen from the initial zero momentum frame, as we will discuss in a future publication.
\par The momentum distribution of the remnant $P_f/m_0$ was evaluated for several $\rho_0$ in the range $0^{o} \leq \rho_0 \leq 90^{}$.
It has the head-on collision case as a lower bound and the orthogonal collision as an upper bound, for all $\alpha$.
A new feature is observed in the final momentum distribution for equal-mass black holes, in the case of non head-on collisions.
Contrary to the case of a head-on collision and of the merging of black hole binary inspirals,
for a non head-on collision of two equal mass black holes the net gravitational
wave flux emitted is nonzero. Therefore the final momentum $P_f$ for $\alpha=1$ is nonzero, for any $\rho_0 \neq 0$,
even if measured in an inertial frame with velocity ${\bf v}_{in}={\bf P}(u_0)/M_B(u_0)$ (the zero initial-Bondi-momentum frame)
relative to a rest inertial frame at infinity. The issue of kick velocities of the merged system due to the recoil of
gravitational wave emission is currently being examined. Our analysis is based
on the impulse conservation equations (\ref{eq33vv}) and we obtain that, in a zero initial Bondi momentum frame,
the velocity of the center-of-mass at $u=u_f$ coincides approximately with the kick velocity.
The distribution of the kick velocity as a function of the symmetric mass parameter satisfies a modified
Fitchett-Blanchet law\cite{fitchett}. This modification is necessary to
give account of the fact that, in a non head-on collision of two equal-mass black holes, the net gravitational wave flux
emitted is nonzero, contrary to the case of a head-on collision and of merging of black hole inspirals.
\par The angular pattern of the gravitational waves emitted (with its two modes of polarization included) is examined
for the case of a low mass ratio parameter and a large incidence angle. In the plane of the collision the pattern
is typically bremsstrahlung, with two dominant lobes in the forward direction of motion of the merged system,
characteristic of a strong initial decelerated regime of the merged system. We show that
gravitational waves are also emitted outside the plane of the collision but that this radiation has a zero net momentum flux.
Therefore it does not extract momentum of the system, in accordance with the planar nature of the collision.

\par Finally we also obtained a relation between the initial incidence angle $\rho_0$ and the scattering angle $\rho_f$
that closely reproduces a result for the inelastic collision of classical particles in Newtonian dynamics, and is
validated by the numerical results. For $\alpha=1$ we recover numerically the exact Newtonian result, with the
$\rho_f=(180^{o}- \rho_0)/2$ being a consequence of the symmetry of the initial data as if the two initial black
holes and the remnant were individual classical particles. This result suggests a Newtonian signature present in the dynamics
of the merged system with initial data (\ref{eq3_13-ii}.
\par The authors acknowledge the partial financial support of CNPq/MCT-Brazil, through a Post-Doctoral Grant No. 201879/2010-7
(RFA), Research Grant No. 306527/2009-0 (IDS), and of FAPES-ES-Brazil (EVT). RFA acknowledges the hospitality of the Center
for Relativistic Astrophysics, Georgia Institute of Technology, Atlanta, GA, USA.


\begin{thebibliography}{99}

\bibitem{rt} I. Robinson and A. Trautman, Phys. Rev. Lett. 4, 431 (1960); Proc. Roy. Soc. A265, 463 (1962).

\bibitem{aranha1} R. F. Aranha, I. Dami\~ao Soares and E. V. Tonini, Phys. Rev. D81, 104005 (2010).

\bibitem{aranha2} R. F. Aranha, I. Dami\~ao Soares and E. V. Tonini, Phys. Rev. D82, 104033 (2010).

\bibitem{pretorius} F. Pretorius, in {\it Physics of Relativistic Objects in Compact Binaries: from Birth to
Coalescence}, edited by M. Colpi, P. Casella, V. Gorini, U. Moschella and A. Possenti (Astrophysics and Space Science Library Series, Vol. 359, Springer, Heidelberg, 2009), p. 305.

\bibitem{boyle} L. Cadonatti et al., Class. Q. Grav. 26, 114008 (2009); M. Boyle, D. A. Brown and L. Pekowsky, Class. Q. Grav. 26, 114006 (2009);
M. Hannam, Class. Q. Grav. 26, 114001 (2009).

\bibitem{AH} K. P. Tod, Class. Quantum Grav. 3, 1169 (1986); K. P. Tod, Class. Q. Grav. 6, 1159 (1989).

\bibitem{bondi} H. Bondi, M. G. J. van der Berg, and A. W. K. Metzner, Proc. R. Soc. London
A 269, 21 (1962).

\bibitem{sachs} R. K. Sachs, Proc. R. Soc. London A 270, 103 (1962); R. K. Sachs, J. Math. Phys. 3, 908 (1962).

\bibitem{winicour} Jeffrey Winicour, \textit{Characteristic Evolution and Matching}, Living Rev. Relativity 8,
(2005),  10. http://www.livingreviews.org/lrr-2005-10.

\bibitem{chrusciel} P. Chrusciel, Commun. Math. Phys. 137, 289 (1991); Proc. Roy.
Soc. London A 436, 299 (1992); P. Chrusciel and D. B. Singleton, Commun.
Math. Phys. 147, 137 (1992).

\bibitem{petrov} A. Z. Petrov, Sci. Nat. Kazan State University 114, 55 (1954); F. A. E. Pirani,
\textit{Introduction to Gravitational Radiation Theory}, in Lectures on General Relativity, Brandeis Summer
Institute in Theoretical Physics, vol. 1 (Prentice-Hall, New Jersey, 1964).

\bibitem{peeling} R. Sachs, Proc. Roy. Soc. A 264, 309 (1961); E. T. Newman and R. Penrose,
J. Math. Phys. 3, 566 (1962).

\bibitem{sachs1} R. K. Sachs, Phys. Rev. 128, 2851 (1962).

\bibitem{york} J.W. York Jr.,
\textit{The initial value problem and dynamics}, in {\textit Gravitational Radiation}, N. Deruelle and T. Piran, Editors, North-Holland (1983).

\bibitem{aranha3} R. F. Aranha, H. P. Oliveira, I. Dami\~ao Soares and E. V. Tonini,
Int. J. Mod. Phys. D 17, 2049 (2008).

\bibitem{arfken} G. Arfken, {\it Mathematical Methods for Physicists}, \S 2.14,
Academic Press (New York, 1968).

\bibitem{stachel} R. A. D'Inverno and J. Stachel, J. Math. Phys. 19, 2447 (1978);
R. A. D'Inverno and J. Smallwood, Phys. Rev. D22, 1233 (1980).

\bibitem{stachel1} J. A. Vickers in {\it Approaches to Numerical Relativity},
ed. R. D'Inverno, Cambridge University Press (Cambridge, 1992).

\bibitem{aranha4} R. F. Aranha, I. Dami\~ao Soares and E. V. Tonini,
submitted to Physical Review D (2011).

\bibitem{saa} R. P. Macedo and A. Saa, Phys. Rev. D 78, 104025 (2008).

\bibitem{korn} G.A. Korn, T.M. Korn, \textit{Mathematical Handbook for Scientists and Engineers},
McGraw-Hill (1967).

\bibitem{unti} E. T. Newman and T. W. J. Unti, J. Math. Phys. 3, 891 (1962).

\bibitem{kramer} U. G\"onna and D. Kramer, Class. Q. Grav. 15, 215 (1998).

\bibitem{eardley} D. Eardley, \textit{Theoretical models for sources of gravitational waves},
in {\it Gravitational Radiation}, N. Deruelle and T. Piran, Editors, North-Holland (1983).


\bibitem{fitchett} M. J. Fitchett Mon. Not. R. Astron. Soc 203, 1049 (1983);
L. Blanchet, M. S. S. Qusailah, and C. M. Will, Astrophys. J. 635, 508 (2005).

\bibitem{penrose} E. T. Newman and R. Penrose, J. Math. Phys. 3, 566 (1962);
S. Chandrasekhar, {\it The Mathematical Theory of Black Holes}, Oxford (1983).

\bibitem{aranhaTese} R. F. Aranha, {\it Gravitational Wave Emission in the Merger of Black Holes: A Theoretical
and Computational Modelling in the Characteristic Formalism}, DSc. Thesis, Centro Brasileiro de Pesquisas F\'isicas,
Rio de Janeiro, May 2011 (unpublished).


\end{thebibliography}
\end{document}